\documentstyle[12pt,epsfig]{report}
\begin{document}
\begin{center}
\null
\vspace{2cm}
{\Large{{\bf{The Antiferromagnetic Sawtooth Lattice -\\ the study of a two spin variant}}}}

\vspace{4cm}

A Project Report\\
Submitted in partial fulfillment of the\\
requirements for the Degree of\\
{\bf{\large{Master of Science}}}\\
\vspace{2cm}
by\\
{\bf{V. Ravi Chandra}}

\vspace{3cm}

{\large{Department of Physics\\ Indian Institute of Science \\ Bangalore - 560012,India}}\\
March, 2003
\end{center}
\pagebreak

\begin{flushleft}
{\large{\bf{ACKNOWLEDGEMENTS}}}\\
\end{flushleft}
{\small
This project wouldn't have been a success and my stay here at IISc wouldn't have
been as fruitful as it has been, without the help and support I have received from
a lot of people around me. This is a good place to remember them and  thank them
for all they have done for me.

Working with Diptiman was a good learning experience and just as importantly, great fun.
I hope that our future projects together are just as good, if not better. Touchwood.

I thank our collaborators Prof.Johannes Richter and Prof.Nedko Ivanov for including me in the
collaboration and allowing me to work on this problem for my MS.

I thank Raghu for many useful discussions on field theory and Sandeep for helping me out
with my computations in my first year. Ritesh is a caring friend we all cherish at CTS.
I thank Srikanth for the most wonderful time I had with him conversing on topics ranging
from  linguistics to horror movies. Siddhartha, Anu, Udit, and Prusty remain dear friends
and I wish all of them well with the new lives that they are building in Germany.

That brings me to the Physics Department where I spent my first one and a half years at IISc.
The list here is so long that I wont risk being specific. I thank my batchmates, seniors
and everybody else that I am aquainted with in that Department for being such wonderful people.
I never had to look too far for anything. Be it that book that just had to be borrowed from someone,
or that printout that just had to be taken at 3:30 am. Or amicable  company to sit and engage in
some plain gossiping and chatting ! Thank you all for everything.

Finally to the extent that such things can be thanked for, I thank my parents and sisters
for all their love, support and patience, the strength and extent of each which I am sure
will be tested to its limits as I go on to spend the coming (few I hope !) years at IISc.

All said and done, I must admit that having written a report like this I don't know whether 
it is fair to have an expectation that somebody will read it.  And I don't mean those
who will read this because they have to ! To any such noble soul 
who perseveres to go beyond this page and tries to get some feeling for what this 
thing is actually about, and appreciates the hard work that has gone in,
I wholeheartedly dedicate this effort.}

\begin{flushleft}
Thanking one and all,\\
Ravi
\end{flushleft}

\pagebreak

\null 
 \vspace{6cm}    
\begin{center}      
{\large{\bf{\underline{ABSTRACT}}}}\\
\end{center}
\begin{flushleft}
Generalising recent studies on the sawtooth lattice, a two-spin variant of the
model is considered. Numerical studies of the energy spectra and the relevant spin
correlations in the problem are presented. Perturbation theory analysis of the model
explaining some of the features of the numerical data is put forward and the spin wave spectra 
of the model corresponding to different phases are investigated.

\end{flushleft}
\pagebreak

\tableofcontents

\chapter{Introduction}
There has been a  lot of recent interest (\cite {skp}, \cite{ivanov1} to \cite{ivanov4})
in one dimensional and quasi-one-dimensional 
quantum spin systems having two different spins in the unit cell with antiferromagnetic couplings.
Depending on the prescence
or the abscence of frustration and its strength when it is present these systems exhibit a rich
variety of phases in the ground state. Quantum  ferrimagnets for example, are one such class of systems
where the system has a finite magnetic moment in the ground state. Chemists have been
successful in synthesising families of organo-metallic compounds (\cite{kahn1} to \cite{gleizes}) 
which provide experimental realisations for some such systems.

In this report we present the study of a mixed spin variant of the {\bf{Sawtooth Lattice}}.
Recent studies (\cite{bss}, \cite{blun}, \cite{rud})
on this model have concentrated on systems where all the spins on the 
lattice are the same. The compound Delafossite ($YCuO_2.5$) provides an experimental
realisation of such a model with the copper ions forming a lattice of spin-$\frac{1}{2}$ sites. 
In an attempt to generalise such studies we considered
a two spin variant of the above model and studied it numerically and analytically. In this
chapter the model and its Hamiltonian are introduced and some of the classically expected
properties of the ground state are discussed. In subsequent chapters the numerical data and 
the analytical results obtained have been presented.

\pagebreak

\section{The model and its classical ground states}
The model under consideration is described by the Hamiltonian,

\begin{equation}
H = J_1{\sum_n  {\overrightarrow S}_{1n}\cdot{\overrightarrow S}_{1(n+1)}}
+ J_2 \sum_n  ({\overrightarrow S}_{2n}\cdot{\overrightarrow S}_{1n} +
{\overrightarrow S}_{2n}\cdot{\overrightarrow S}_{1(n+1)})
\end{equation}
Or equivalently,
\begin{equation}
H = \delta {\sum_n  {\overrightarrow S}_{1n}\cdot{\overrightarrow S}_{1(n+1)}}
+ \sum_n  ({\overrightarrow S}_{2n}\cdot{\overrightarrow S}_{1n} +
{\overrightarrow S}_{2n}\cdot{\overrightarrow S}_{1(n+1)})
\end{equation}
where $\delta \equiv {J_1}/{J_2}$ and $J_2$ has been set to 1. Thus all energies in 
the problem are measured in units of $J_2$. Here $S_{1i}$ denotes a spin-1 site and $S_{2i}$ denotes
a spin-$\frac{1}{2}$.\\
Schematically the model looks like,

\begin{flushleft}
\epsfxsize=16cm \epsfysize=3.0cm \epsfbox{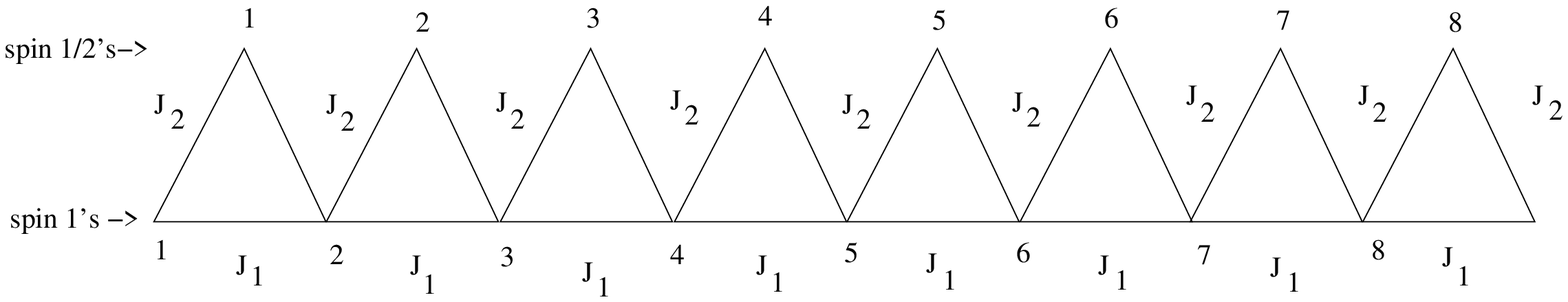}
\end{flushleft}

Classically, the ground state of this system is characterised by a planar
or a collinear arrangement of spin vectors. Which of these arrangements is the ground state 
depends on the relative strengths of the two couplings $J_1$ and $J_2$.

When $\boldmath J_2S_2>2J_1S_1$ the classical ground state is characterised
by a collinear arrangement of spins and this phase is called the 
{\bf{ferrimagnetic state}}. Schematically the ferrimagnetic phase looks as follows:

\begin{flushleft}
\epsfxsize=16cm \epsfysize=2cm \epsfbox{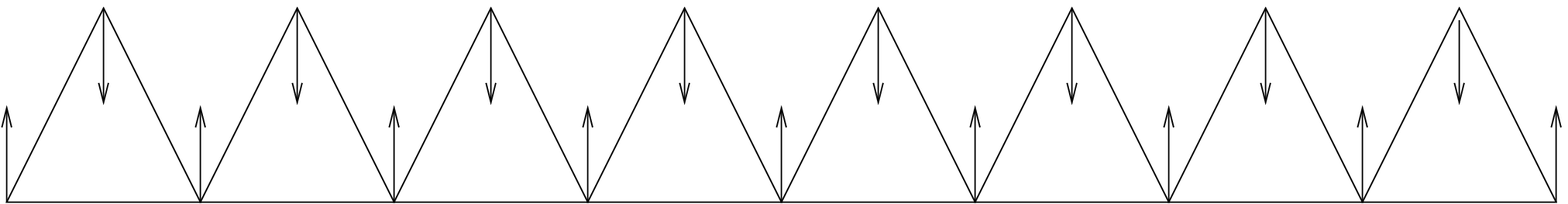}
\end{flushleft}
\pagebreak
When  $\boldmath J_2S_2<2J_1S_1$, the classical ground  state is characterised
by a planar arrangement of spins in each triangle and this phase is called the {\bf spiral/canted}
phase. Schematically this phase looks as follows:

\begin{flushleft}
\epsfxsize=15cm \epsfysize=4cm \epsfbox{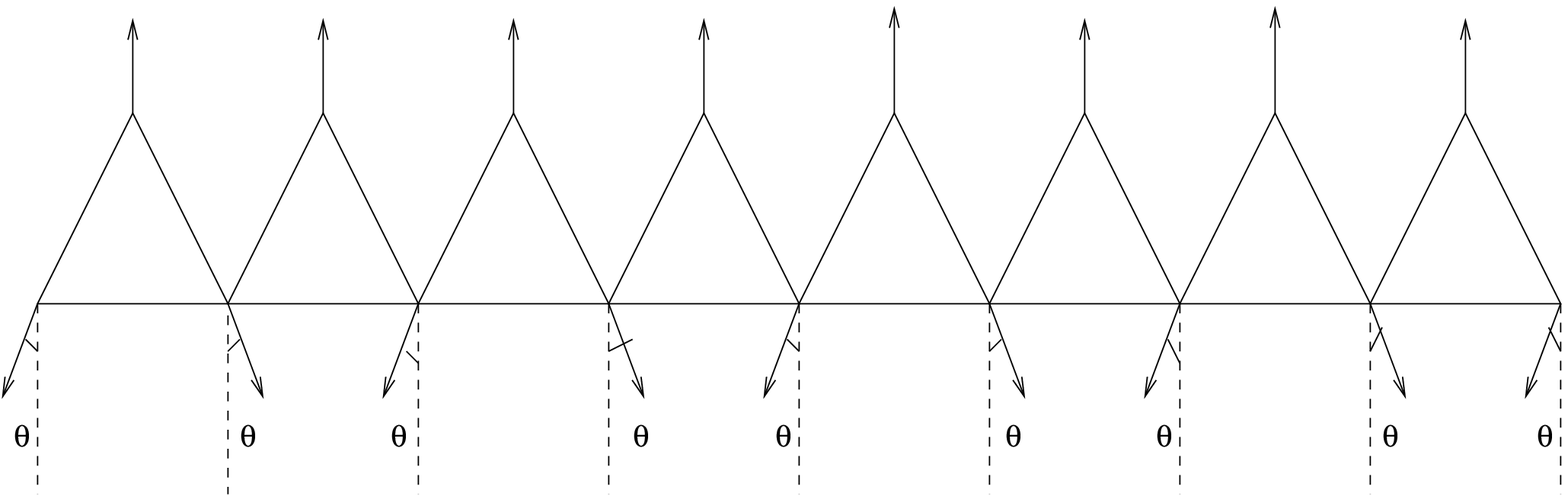}
\end{flushleft}
where $\cos \theta \equiv {J_2S_2}/{2J_1S_1}$. Because of the freedom of
choosing the direction of spins even when the spin vectors in each triangle 
are constrained to be on a plane, the classical spiral/canted phase has an infinite 
amount of degeneracy. We note that because of the above reason though the spin vectors
in a particular triangle have to be in a plane, all the spin vectors need not lie in the same plane. 

For our simulations and the analytical results that follow, the parameter in the problem
is $\delta$. Numerically we have studied the energy spectra and the spin correlations 
as a function of $\delta$. These results are presented in the next chapter. And perturbation
theory calculations to compute the effective Hamiltonian between
the spin-{$\frac{1}{2}$}'s in the large $\delta$ limit and the spin wave spectra 
obtained for the above two phases are presented in chapters 3 and 4 respectively.

\chapter{Exact Diagonalization and related results}
Numerical studies of the model have been done using the Exact Diagonalisation method.
By calculating the required eigenvalues and eigenvectors using the Lanczos algorithm
the following quantities were calculated:
\begin{itemize}
\item{The variation of the ground state and the first excited state energies with $\delta$.}
\item{The correlations between the spins in the ground state.}
\item{The effective Hamiltonian governing the spin-{$\frac{1}{2}$}'s when the coupling 
between the spin-1's is much stronger than $J_2$, the  spin-1 - spin-{$\frac{1}{2}$} coupling .}
\end{itemize} 

In this chapter we begin with a brief introduction to the Lanczos algorithm and the 
version of it which has been used. Following that the numerical results of the first two 
categories above are presented. The effective Hamiltonian calculations, being semi-analytical 
in nature are presented in a subsequent chapter.

\section{The Lanczos Algorithm}
The {\bf{Lanczos Algorithm}} is a widely used method for finding a few eigenvalues
of a large symmetric matrix. Since the matrices that one deals with in quantum spin 
systems are usually symmetric (or Hermitian,whose eigenvalue problem can be formulated as one
of a symmetric matrix double the size), this method is commonly 
used to find the lowest few eigenvalues
and eigenvectors corresponding to the ground state and the lowest excited states.\\

The basic content of the version of the Lanczos procedure used for our simulations
is the following recursion relation.

\begin{equation}
\beta_{i+1}v_{i+1} = Av_{i} - \alpha_{i}v_{i} - \beta_{i}v_{i-1}
\end{equation}
for i=1,2 .....  \\  
where A is the matrix whose eigenvalues we want to calculate and  
$\alpha_i \equiv {v^{T}}_{\hspace{-0.3cm}i} A v_i  $ and 
$\beta_{i+1} \equiv {v_{i+1}}^{{\hspace{-0.5cm} T}} \hspace{0.3cm} A  v_i$. 
$\beta_1$ is taken to be 0 and $v_1$ is chosen to be a random vector
normalised to unity.
For any $i=m$ the a symmetric tridiagonal matrix $T_m$ is defined whose diagonal elements are $\alpha_i$
and the off-diagonal elements are $\beta_j$ (j=2,m). It can be proved (for infinite precision
arithmetic) that if $\lambda_1>=\lambda_2>=\lambda_3>=......\lambda_m$ are eigenvalues of $T_m$
and  $\Lambda_1>=\Lambda_2>=\Lambda_3>=......\Lambda_m$ are the m largest eigenvalues of $A$,
then the sequence of $\lambda_i$'s converges to the sequence of $\Lambda_i$'s as m is incremented.

Again, let $V_m$ be the $n \times m$ (where n is the order of A) whose $i$th column
is $v_i$. Then if $x_i$ is the eigenvector of $\lambda_i$ and $X_i$ is the eigenvector 
of $\Lambda_i$ then $V_m x_i \longrightarrow X_i$ as m is incremented. The vectors
$V_m x_i$ are called {\bf{Ritz vectors}}.  

The Lanczos vectors $v_i$ which are generated by the above recursion are an 
orthonormal set. The proof of the two claims made above hinges on this orthonormality
of the Lanczos vectors. In reality when these vectors are generated numerically,
finite precision effects enter and the set of vectors generated are not strictly
orthogonal. The loss of orthogonality of the Lanczos vectors affects the computations 
primarily in two ways. Eigenvalues which are simple appear as multiple eigenvalues
of the system and more importantly spurious eigenvalues which are not eigenvalues 
of $A$ at all appear as eigenvalues of $T_m$. 

Various methods have been devised to overcome the difficulties posed by 
this loss of orthogonality. One way is to resort to reorthogonalisation
of the Lanczos vectors. We don't use this method in our computations.
We instead use the identification test developed by Cullum and Willoughby (\cite{cullum})
to explicitly identify the eigenvalues which are spurious and discard them as
they are detected.

As already mentioned in finite precision calculation the appearance of an eigenvalue 
as a multiple eigenvalue of the tridiagonal matrix $T_m$ is no guarantee of that eigenvalue
being a true multiple eigenvalue of $A$. But this difficulty can be overcome by looking at the
corresponding Ritz vectors. If Ritz vectors are calculated for a large enough $T_m$ for which 
a particular eigenvalue is duplicate then for  a true multiple eigenvalue two 
linearly independent Ritz vectors can be generated using appropriate $T_m$'s of
different sizes. 
But if the eigenvalue is simple any two Ritz vectors of the same eigenvalue will 
essentially be the same (upto a sign). In this way we can determine the degeneracy of eigenvalues 
by computing more and more Ritz vectors and checking for linear independence. In our program we
employ this method to determine the degeneracy of an eigenvalue.

\section{Results of Exact Diagonalisation}

All the  computations have been done by generating the Hamiltonian in the total $S_z$ basis.
The variation of various energies and correlations with the ratio of interaction strength
$\delta = {J_1}/{J_2}$ has been studied. In all the results reported $\delta$ varies from
0.1 to 2. The correlations have been calculated for a system size of 10 triangles and 
all the other graphs are for a system size of 8 triangles. All through the computations periodic
boundary conditions have been used (the spin-1's are joined in a ring).
For all the reported data 
the accuracy measured by $\vert{{{(AX - \Lambda X)}^{T}{(AX - \Lambda X)}}}\vert$ for an 
eigenvalue $\Lambda$ and its corresponding eigenvector $X$ (normalised to unity) 
is $\sim 10^{-12}$  or less.

\pagebreak

\subsection{Variation of ground state energy and first excited state energy with $\delta$}

\begin{center}
\epsfxsize=9cm \epsfysize=9cm \epsfbox{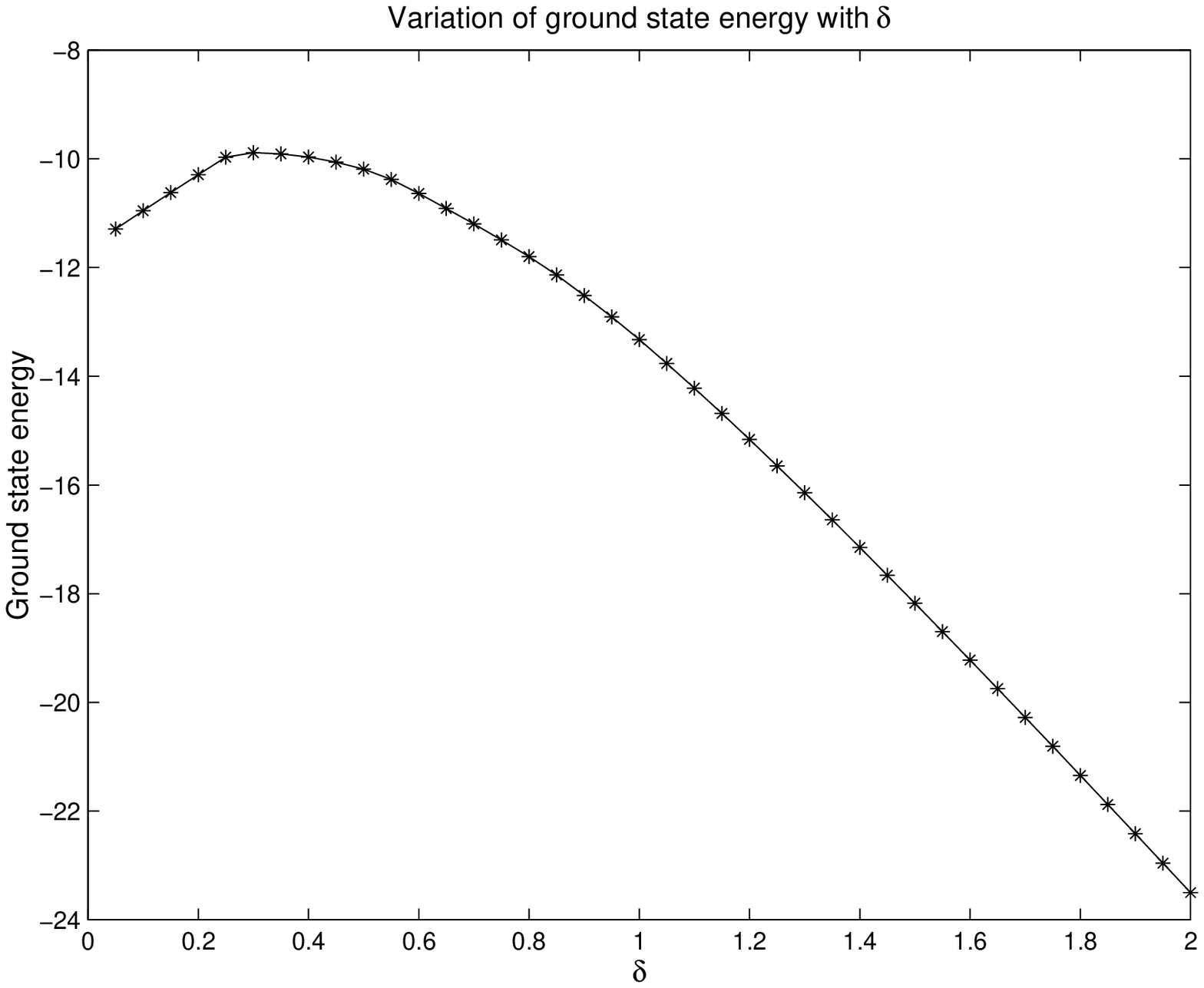}
\end{center}

\begin{center}
\epsfxsize=9cm \epsfysize=9cm \epsfbox{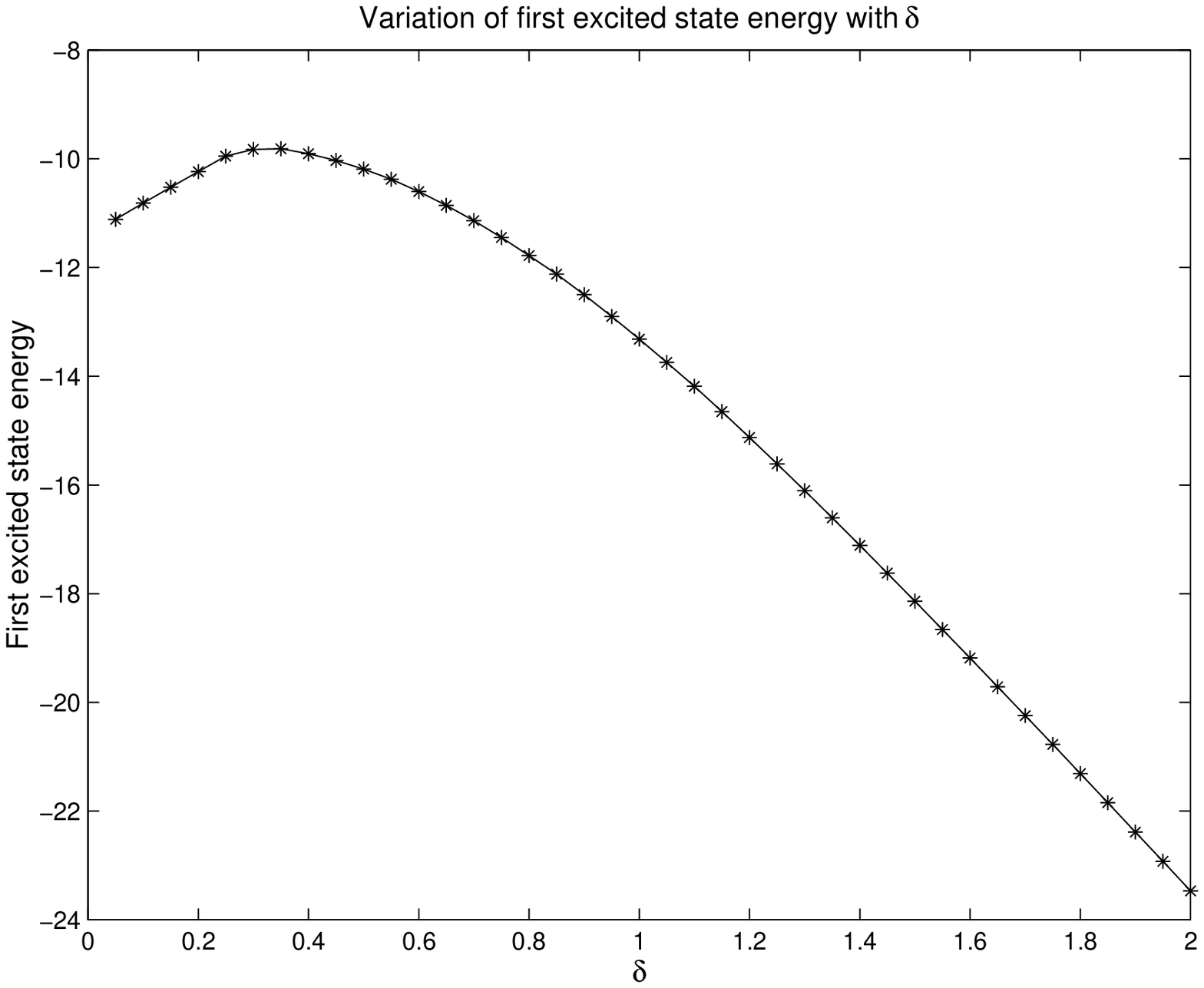}
\end{center}

\subsection{Variation of total spin of the ground and first excited states with $\delta$}

\begin{center}
\epsfxsize=9cm \epsfysize=9cm \epsfbox{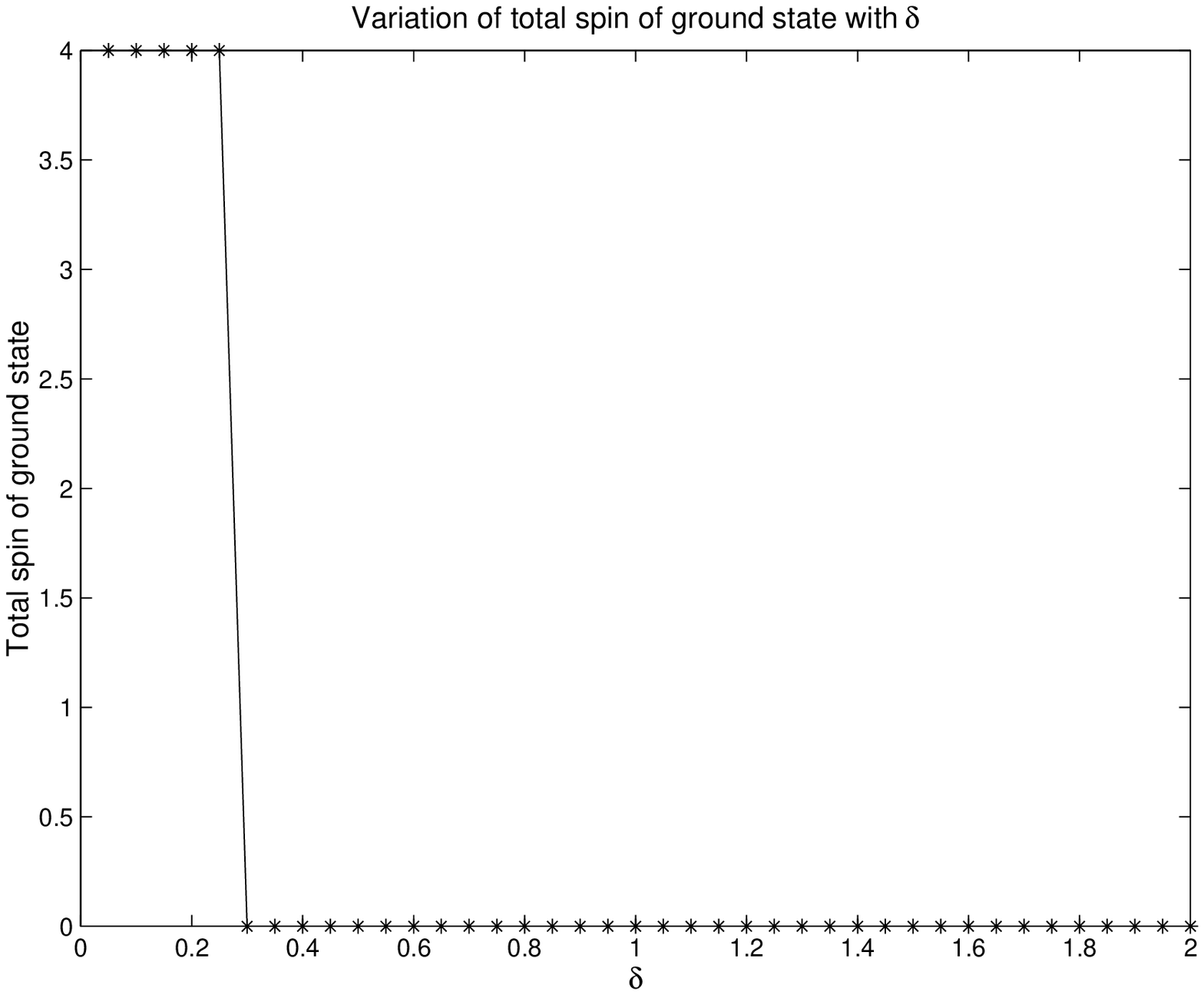}
\end{center}

\begin{center}
\epsfxsize=9cm \epsfysize=9cm \epsfbox{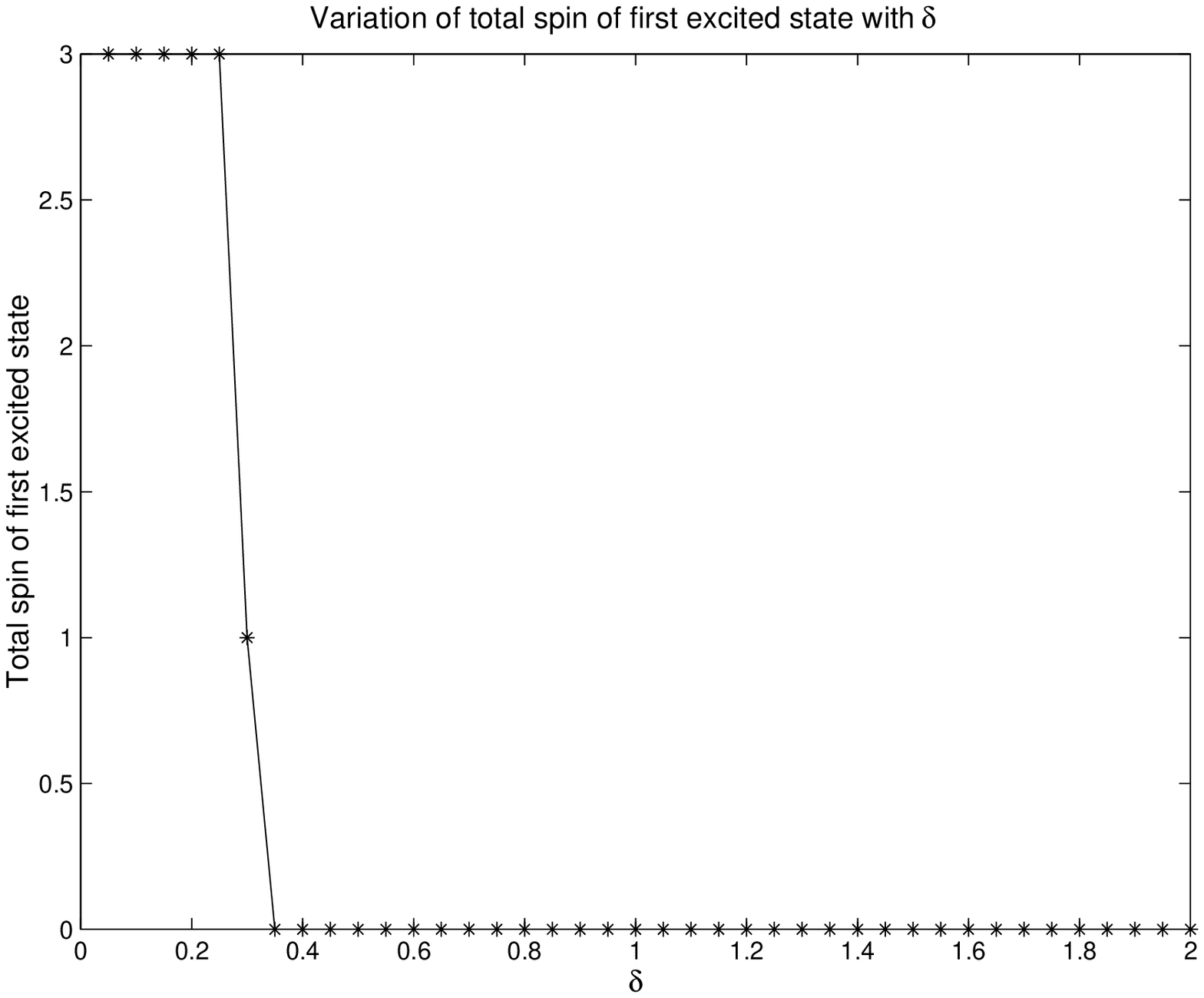}
\end{center}

\subsection{Variation of energy gap to the 1st excited state with $\delta$} 

\begin{center}
\epsfxsize=10cm \epsfysize=10cm \epsfbox{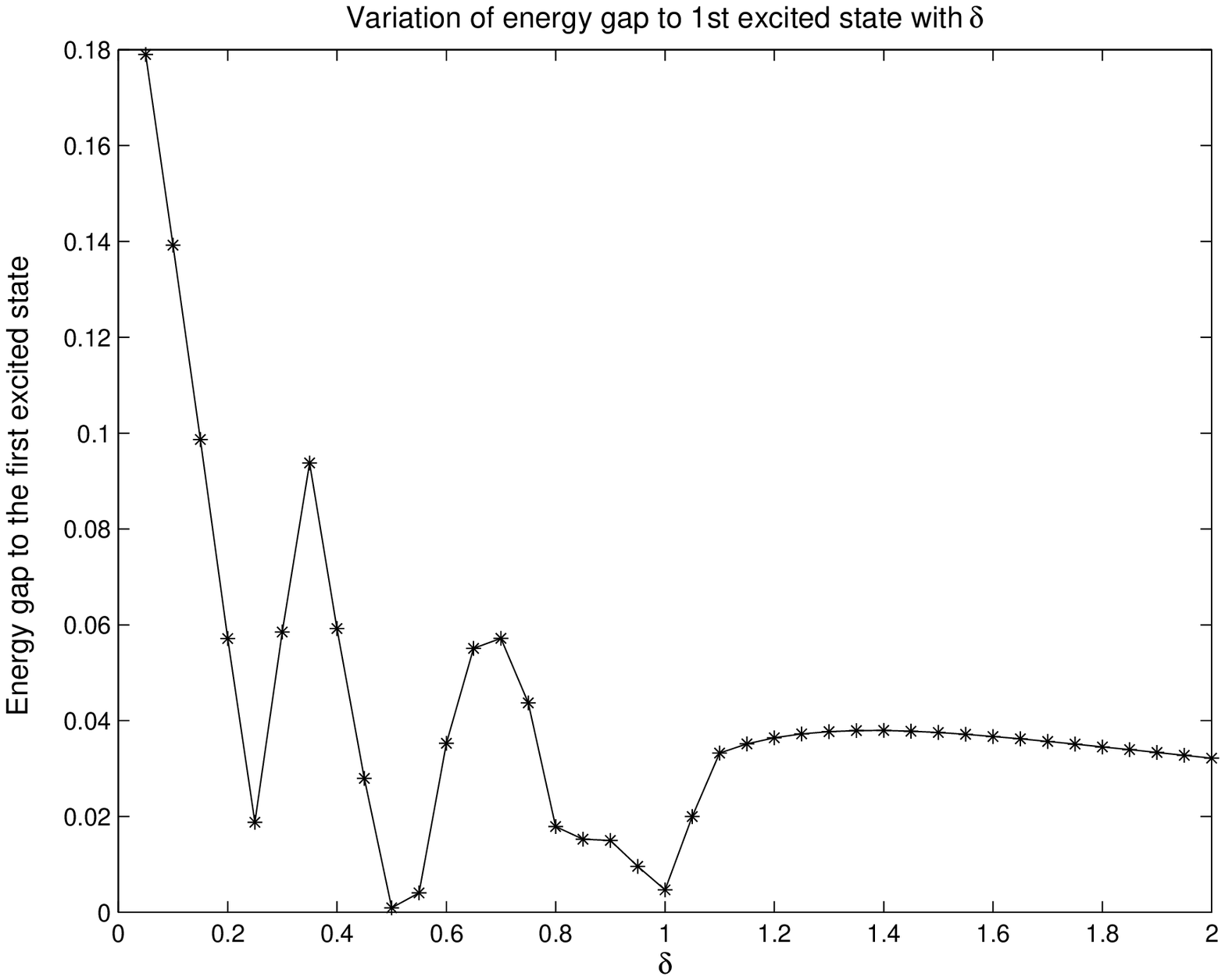}
\end{center}

Significant aspects of the above results are the following:

\begin{itemize}
\item{For a small $\delta$ this sawtooth lattice can be approximated by an
alternating spin-1/spin-$\frac{1}{2}$ chain. This system has a ferrimagnetic 
ground state of total spin $N(S_1 - S_2)$ and the first excited
state is a state of total spin one less than that of the ground state (\cite{skp}). One can see
clearly from the figures that for small $\delta$ this is indeed the case for this system.}

\item{There is a sudden change in the total spin of both the ground state and the first
excited state at around $\delta = 0.25$. This is the point where we expect the
transition from ferrimagnetic to the spiral phase from the classical analysis. We have 
analysed this particular region closely by studying the total spin behaviour of the 
ground state at a number of closely spaced points from $\delta=0.2$ to $\delta=0.35$.
Numerically we have found that for the ground state the spin drops to zero at $\delta = 0.265$.}

\item{There are two values of $\delta$ where the system seems to be gapless. The first is near
$\delta = 0.5$ and the other is near $\delta = 1.0$. This actually divides the phase diagram
into three phases as opposed to the two phases that we expected classically. The nature of the
two quantum phases other than the ferrimagnetic phase is not clear as of now.} 

\item{For any  $\delta$ if all the spin-$\frac{1}{2}$
interactions are made zero($J_2=0$) then the ground state
energy must essentially be that of a spin-1 Heisenberg antiferromagnet. In the
thermodynamic limit this energy per site has been
calculated {\cite{huse}} to very good
accuracy.
Our result ($E_0/N = -1.41712J_1$) for the 8 site cluster agrees  
with the above value ($E_0/N =-1.40148J_1$)  upto finite size effects.}
\end{itemize}

\section{Spin correlations in the ground state}

All the spin correlation calculations have been done with a system size of 10 triangles.
In order to check the accuracy of the data the following checks were employed. 

\begin{itemize}

\item{The three correlations $<S_{i}^zS_{j}^z>$,
$<{S_{i}}^{+}{S_{j}}^{-}>$,$<{S_{i}}^{-}{S_{j}}^{+}>$ were calculated separately.
Then for each case it was checked that the latter two were equal and for $\delta$'s for which 
the eigenstate was a singlet it was checked that the $<\overrightarrow S_{i}.\overrightarrow S_{j}>$
(which can be computed from the above three was thrice the  $<S_{i}^zS_{j}^z>$ correlation as 
would be expected for a singlet.}

\item{For the singlet states it was verified that $\sum_{j} <\overrightarrow S_{i}.\overrightarrow S_{j}>$
where j runs over all the spins for a given i was zero.}

\end{itemize}

In the graphs that follow, the  $<S_{i}^zS_{j}^z>$ correlations are reported. The numbering scheme for 
the spins is same as shown in the figure in chapter 1.

\subsection{Spin-1 - Spin-1 correlations in the ground state}

\begin{center}
\epsfxsize=9cm \epsfysize=9cm \epsfbox{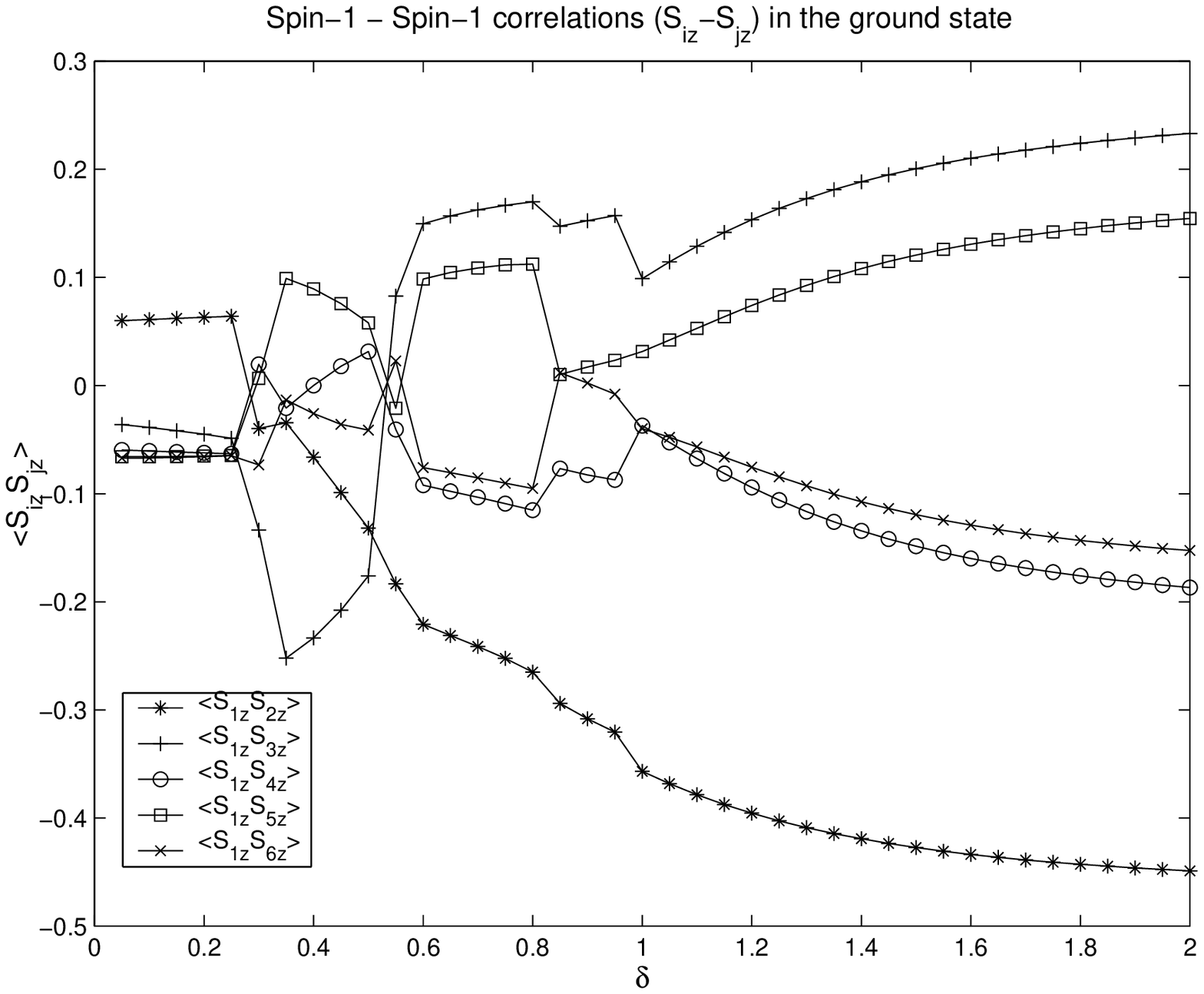}
\end{center}

\subsection{Spin-1 - Spin-$\frac{1}{2}$ correlations in the ground state}
\begin{center}
\epsfxsize=9cm \epsfysize=9cm \epsfbox{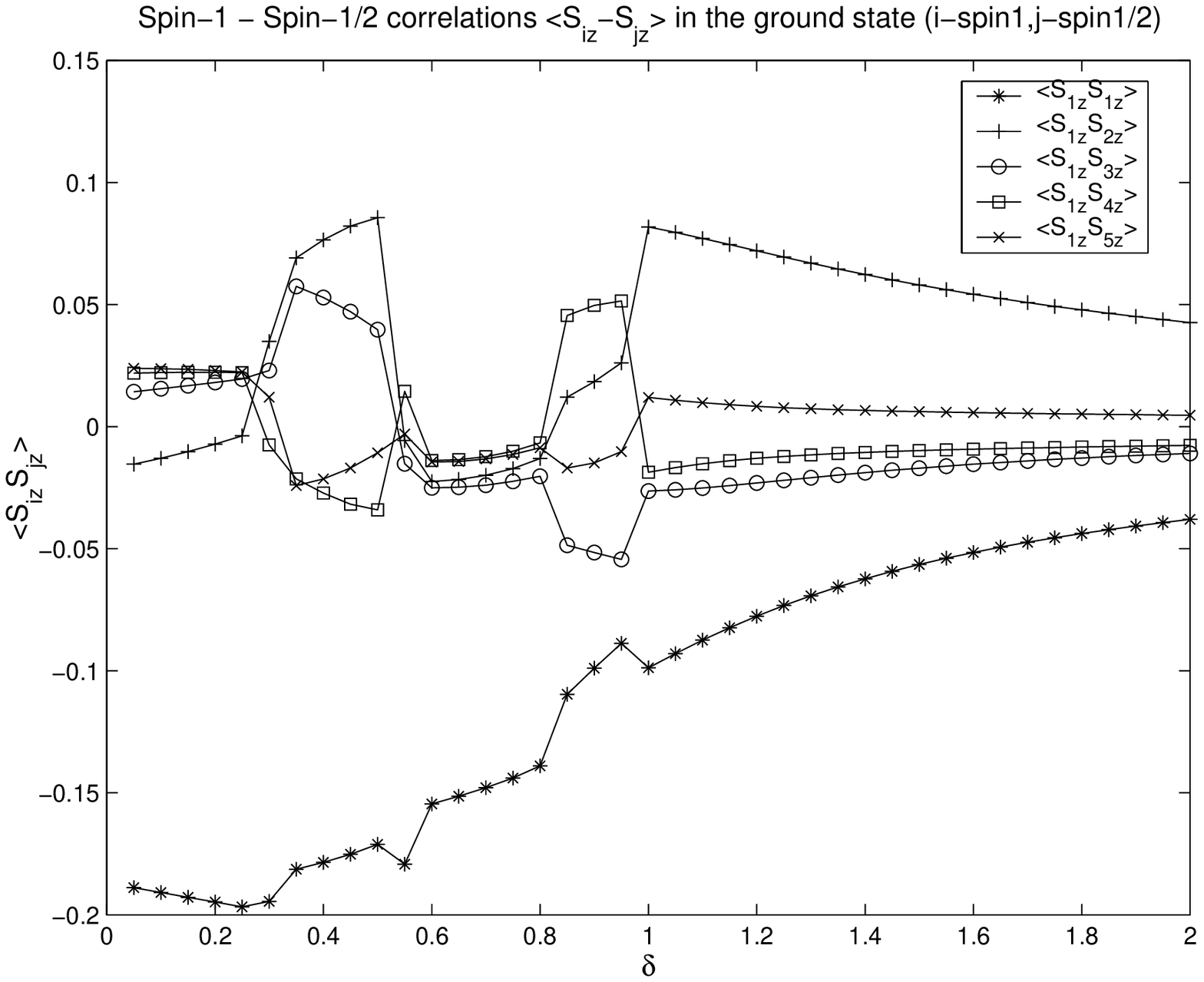}
\end{center}

\subsection{Spin-$\frac{1}{2}$ - Spin-$\frac{1}{2}$ correlations in the ground state}
\begin{center}
\epsfxsize=10cm \epsfysize=10cm \epsfbox{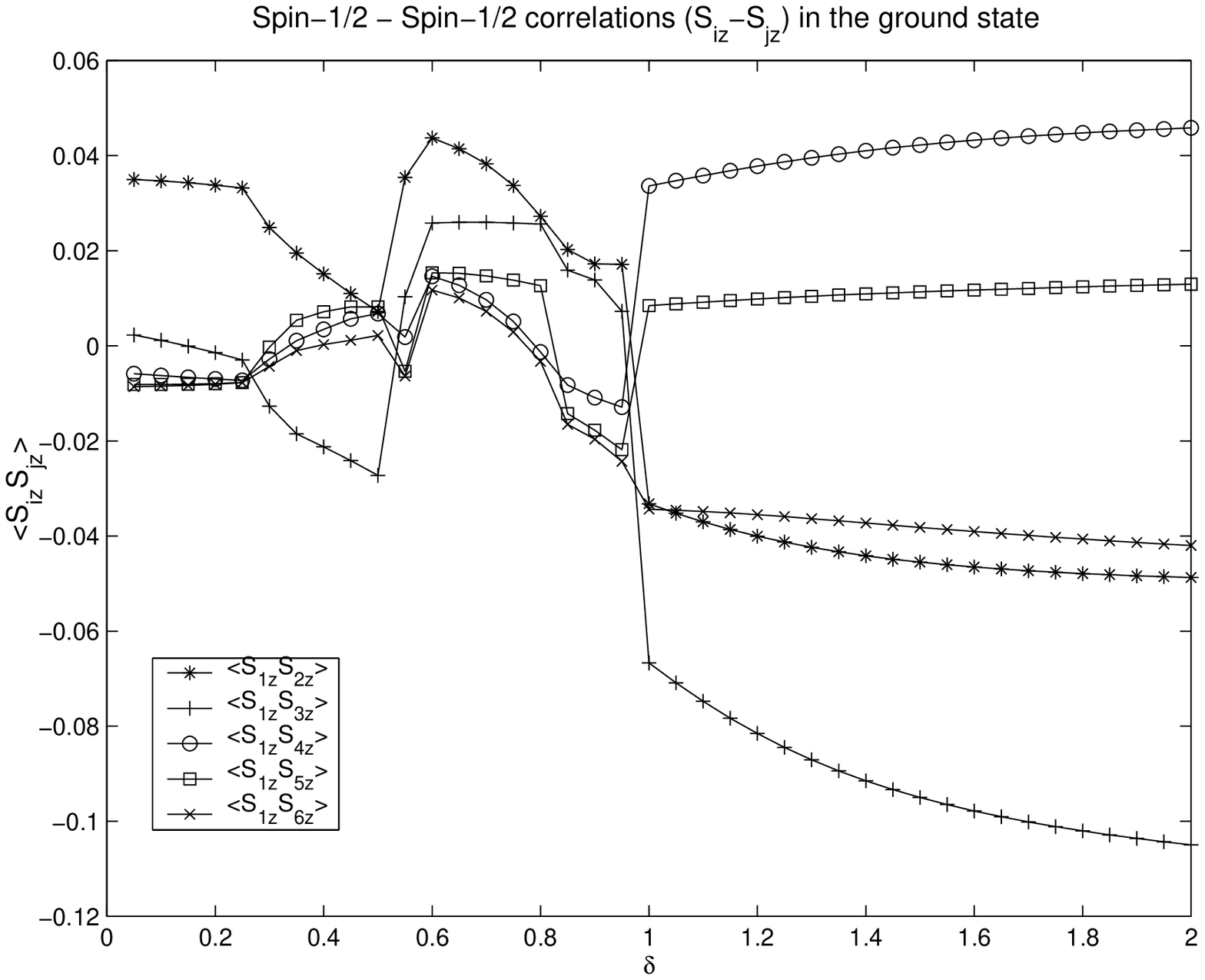}
\end{center}

Salient features of the correlation function plots are the following: 

\begin{itemize}
\item{Though for most of the region the correlation functions behave regularly, in the neighbourhood of 
$\delta=0.5$ and $\delta=1.0$ the correlations vary very rapidly with $\delta$. From a  preliminary examination 
of the data it seems  that the rate of change of correlations  with $\delta$ is  divergent at $\delta=1.0$ .}

\item{The spin-1 - spin-1 correlations show the expected behaviour as we go towards the large $\delta$ limit.
The spin correlations are alternating in sign as we move from one spin to the next (going farther from
the first spin) and also decaying with distance as we would expect for a pure spin-1 system.}

\item{The spin-1 - spin{$\frac{1}{2}$} correlations all tend to zero in the large $\delta$ limit as we see from
the plot. This is to be expected as in the large $\delta$ limit the system behaves essentially as a spin-1
system which forms a singlet and all the orderings of spin halves are essentially degenerate
as their interaction is much weaker in comparison to the spin-1 - spin-1 interaction. 
Thus the spin-1's and spin-$\frac{1}{2}$'s are not strongly correlated and thus the correlations
tend to zero.}

\item{The most interesting behaviour in the large $\delta$ limit is shown by the spin-{$\frac{1}{2}$} - spin-{$\frac{1}{2}$}
correlations. The most striking feature in this plot is that the next-nearest-neighbour correlation is larger than
all the other correlations, even the nearest neighbour one. This and some of the other features can be 
explained if we calculate the effective Hamiltonian governing the spin-{$\frac{1}{2}$}'s 
in the  large $\delta$ limit.
This calculation and the results that come forth from that analysis have been presented in chapter 3.} 

\end{itemize}

\chapter{Perturbation theory and the effective Hamiltonian}
The model Hamiltonian we have been concerned with is the following:

\begin{equation}
H = J_1{\sum_n  {\overrightarrow S}_{1n}.{\overrightarrow S}_{1(n+1)}}
+ J_2 \sum_n  ({\overrightarrow S}_{2n}.{\overrightarrow S}_{1n} +
{\overrightarrow S}_{2n}.{\overrightarrow S}_{1(n+1)})
\end{equation}
where $S_1$'s are the spin-1 sites and $S_2$ are the spin-$\frac{1}{2}$ sites.

In the regime where the interaction between spin-1's is much stronger than the
interaction between the  spin-$\frac{1}{2}$'s, we can consider the second term in the above 
Hamiltonian to be a perturbation to the spin-1 system. By doing perturbation theory 
calculations we can then find the {\bf{effective Hamiltonian}} (for a few low lying states)
governing the  spin-$\frac{1}{2}$'s
once the spin-1's are essentially decoupled from them as a pure spin-1 system.

We have found that to second order in perturbation theory this effective Hamiltonian has 
a particularly simple form which has only two spin interactions involving terms of the form 
$\overrightarrow S_i.\overrightarrow S_j$. In this chapter we set up the formalism to 
find that Hamiltonian and give a proof of the fact that to second order it has the form  
mentioned above. Then we describe a numerical technique to calculate the various interaction
strengths in the problem.

\section{The calculation of effective Hamiltonian}

We calculate the effective Hamiltonian  in the following manner:
\begin{itemize}
\item{The first term of the Hamiltonian in equation 3.1 is treated as the unperturbed 
Hamiltonian $H_0$. And the second term is the perturbation $V$. Let $|\psi_i>$ be the 
eigenstates of $H_0$, the spin-1 system. We assume that $|\psi_i>$ are simultaneous eigenstates of
$H_0$, the total angular momentum ($({\sum_i {\overrightarrow S_{1i}}})^2$)
and the total $S_z (\sum_i S_{1i}^{z})$ operators. Such states can always be found as they form 
a mutually commuting set of operators.} 

\item{The ground state of $H_0$ is known to be a singlet and furthermore it is 
non-degenerate. We calculate the "corrections" to this ground state energy using 
non degenerate perturbation theory.}

\item{The perturbation term $V$ contains both spin-1 and spin-${\frac{1}{2}}$ operators.
But in calculating the corrections to the ground state energy, the  required matrix elements
will be evaluated using the spin-1 system eigenstates. Thus we will be left at 
every order in perturbation theory with spin-${\frac{1}{2}}$ operators and 
their products. Just as we would have got  the perturbative corrections to the 
unperturbed energy eigenvalues of $H_0$
if the above mentioned matrix elements had been numbers, 
here we get perturbative corrections to the
spin-1 Hamiltonian. These corrections order by order 
will constitute the effective Hamiltonian
of the spin-${\frac{1}{2}}$ system. We note here that this effective Hamiltonian
can be used to find only the states of the full Hamiltonian which lie close to the
singlet ground state. That is because it is calculated by evaluating the effect
of the perturbation only on the singlet ground state.}
\end{itemize}
Having set up the formalism we now proceed to calculate the effective 
Hamiltonian.

\pagebreak

\subsection{The effective Hamiltonian: First order}
The effective Hamiltonian to the first order (in $J_2$ ) is given by,

\begin{equation}
\Delta H_1 = J_2 <0| (
 \sum_n  ({\overrightarrow S}_{2n}\cdot{\overrightarrow S}_{1n} +
{\overrightarrow S}_{2n}\cdot{\overrightarrow S}_{1n+1})) |0>
\end{equation}
where we denote by $|0>$ the singlet ground state of the spin-1 system.

Clearly, this is zero. That is because the state $|0>$ is a spherically symmetric
state and the spin-1 operators occur in the above expression linearly. This can also
be argued from Wigner-Eckart theorem. We know that all the spin-1 operators can be
expressed as linear combinations of spherical tensors of rank 1. But the state with 
respect to which the expectation value is being taken is a singlet. Since we cannot add
$J_1=0$ and $J_2=1$ to give $J_{total}=0$, the above expression must be zero.\\

{\bf{So to first order in $J_2$, the effective Hamiltonian vanishes.}}

\subsection{The effective Hamiltonian: Second order}
The effective Hamiltonian to second order in $J_2$ is given by,
\begin{eqnarray}
\hspace{-1.0cm} \Delta H_2  =  &  & \nonumber \\
&\hspace{-0.75cm}  \sum_{k \neq 0} \frac{<0| J_2 {\sum_{n}
({\overrightarrow S}_{2n}\cdot{\overrightarrow S}_{1n} +
{\overrightarrow S}_{2n}\cdot{\overrightarrow S}_{1(n+1)})} |\psi_k><\psi_k| J_2 \sum_{n \prime }
({\overrightarrow S}_{2n\prime }\cdot{\overrightarrow S}_{1n \prime } +
{\overrightarrow S}_{2n\prime }\cdot{\overrightarrow S}_{1(n\prime  +1)})|0>}{E_0 - E_k}   & 
\end{eqnarray}
where the $k\neq0$ implies that the sum is taken over all eigenstates except 
the singlet ground state. $E_i$ is the energy of the state $\psi_i$ and $E_0$
is the energy of the singlet ground state.
\pagebreak

\begin{flushleft}
Let us consider the matrix element,
\begin{center}
    ${<0|({\overrightarrow S}_{2n}\cdot{\overrightarrow S}_{1n} +
{\overrightarrow S}_{2n}\cdot{\overrightarrow S}_{1(n+1)})} |\psi_k>$
\end{center}
\end{flushleft}
The state on the left is a singlet ($\mathbf J=0$) and components of 
both $\overrightarrow S_{1n}$ and $\overrightarrow S_{1(n+1)}$ can be expressed 
as spherical tensors of rank 1. Thus Wigner-Eckart theorem guarantees that the
only those states $|\psi_k>$ will contribute which have a value of 
total angular momentum which when added to $J=1$ can give us $J_{tot}=0$. But that means
that the only allowed value is 1. Thus we come to the conclusion that {\bf{in Eq. 3.3
we only have to sum over such $|\psi_k>$ which are spin 1 states}}.
Equipped with  this simplification we now look at one particular term in eq 3.3
corresponding to a particular spin 1 state. It will look like,

\begin{center}
\large{
$\sum_{i} {\frac{<0| J_2 {\sum_{n}
({\overrightarrow S}_{2n}\cdot{\overrightarrow S}_{1n} +
{\overrightarrow S}_{2n}\cdot{\overrightarrow S}_{1(n+1)})} |\psi_i^{k}><\psi_i^{k}| J_2 \sum_{n ^\prime }
({\overrightarrow S}_{2n^\prime }\cdot{\overrightarrow S}_{1n^ \prime } +
{\overrightarrow S}_{2n^\prime }\cdot{\overrightarrow S}_{1(n^\prime  +1)})|0>}{E_0 - E_k}}$}
\end{center}
where $k$ labels the particular spin 1 state and the sum is over $i$ which labels the 
particular $S_z$ component (i=-1 , 0, 1).

\begin{flushleft}
One generic term in the above sum will look like
\end{flushleft}
\begin{center}
\large{
$J_2^2\sum_{i} {\frac{<0|  {
\overrightarrow S}_{2n}\cdot{\overrightarrow S}_{1n}
 |\psi_i^{k}><\psi_i^{k}|  
{\overrightarrow S}_{2n^\prime }\cdot{\overrightarrow S}_{1n^ \prime } 
|0>}{E_0 - E_k}}$}
\end{center}
We note here the the denominator will be the same for all the $i$'s as the states
have the same total angular momentum. Furthermore, we as of now don't make any assumptions
as to the relative values of $n$ and $n^\prime$. Whatever we derive below will be true 
whether they are equal or not.
In terms of components the numerator of the above expression
will look like,
\begin{center}
{\large{
$\sum_{\alpha \beta} S_{2n}^{\alpha} S_{2n^\prime}^{\beta} {<S_{1n}^{\alpha}>}_{0i}{<S_{1n^\prime}^{\beta}>}_{i0}$}}
\end{center}
where $\alpha$ , $\beta$ = x, y, z.
And ${<S_{1n}^{\alpha}>}_{0i}   \equiv   <0|S_{1n}^{\alpha}|\psi_i>$ and ${<S_{1n}^{\alpha}>}_{i0}$ is the 
complex conjugate of the same (for now we drop the superscript $k$ as we will talk about a particular
spin-1 state).

\pagebreak
We make the following two claims:

\begin{flushleft}
\begin{itemize}
\item{{\boldmath{$\sum_{i} {<{S_{1n}^{x}}>}_{0i} {<{S_{1n^\prime}^{x}}>}_{i0} = \sum_{i} {<{S_{1n}^{y}}>}_{0i} {<{S_{1n^\prime}^{y}}>}_{i0} = 
\sum_{i} {<{S_{1n}^{z}}>}_{0i} {<{S_{1n^\prime}^{z}}>}_{i0}$ \hspace{3cm}  .........  {\bf(A)}}}}
\item{{\boldmath$\sum_{i} {<{S_{1n}^{\alpha}}>}_{0i} {<{S_{1n^\prime}^{\beta}}>}_{i0} = 0$} if {\boldmath$\alpha \neq \beta$} 
\hspace{.38cm} ......... {\bf(B)}}
\end{itemize}

\end{flushleft}
Before proceeding to prove the above we define the following:
\begin{flushleft}
$\sqrt 2 U_{+1} \equiv -S_{1n}^{+}$, $ \sqrt 2 V_{+1} \equiv -S_{1n^\prime}^{+}$, 
$\sqrt 2 U_{-1} \equiv S_{1n}^{-}$, $ \sqrt 2 V_{-1} \equiv S_{1n^\prime}^{-}$, 
$U_{0} \equiv S_{1n}^{z}$, $V_{0} \equiv S_{1n^\prime}^{z}$ 
\end{flushleft}
$U_{\pm 1,0}$ and $V_{\pm 1,0}$ are by definition  components of spherical tensors of rank 1 and
${S_{1n}^{\alpha}}$ and ${S_{1n^{\prime}}^{\alpha}}$ can be expressed as linear
combinations of components of $\mathbf U$ and $\mathbf V$  defined above.\\
We now prove the above two assertions: \\

\begin{flushleft}
{\bf{\underline{Proof of (A)}}}
\end{flushleft}
We consider the x-x term first.
\begin{eqnarray}
{\sum_{i} {<{S_{1n}^{x}}>}_{0i} {<{S_{1n^\prime}^{x}}>}_{i0}} = & \hspace{-2.75cm} \sum_i <0|{S_{1n}^{x}}|\psi_i><\psi_i|{S_{1n}^{x}}|0> &
 \nonumber \\
& \hspace{-1.05cm} =  \sum_{i}<0|\frac{-U_{+1} + U_{-1}}{\sqrt 2}|\psi_i><\psi_i|{\frac{-V_{+1} + V_{-1}}{\sqrt 2}}|0> &
\nonumber \\
& \hspace{-1.48cm} = \hspace{0.15cm} <0|\frac{-U_{+1} + U_{-1}}{\sqrt 2}|\psi_1><\psi_1|{\frac{-V_{+1} + V_{-1}}{\sqrt 2}}|0> &
\nonumber \\
&\hspace{-0.75cm} +  <0|\frac{-U_{+1} + U_{-1}}{\sqrt 2}|\psi_0><\psi_0|{\frac{-V_{+1} + V_{-1}}{\sqrt 2}}|0> &
\nonumber \\
& \hspace{-0.25cm} +  <0|\frac{-U_{+1} + U_{-1}}{\sqrt 2}|\psi_{-1}><\psi_{-1}|{\frac{-V_{+1} + V_{-1}}{\sqrt 2}}|0> &
\end{eqnarray}
where we have summed over the three values of $S_{z}^{tot}$ for the spin-1 state.
\pagebreak

We can now use the m-selection rule to eliminate those terms above which are zero. After doing that we see that
the above expression reduces to:

\begin{eqnarray}
{\sum_{i} {<{S_{1n}^{x}}>}_{0i} {<{S_{1n^\prime}^{x}}>}_{i0}} = & 
   <0|\frac{ U_{-1}}{\sqrt 2}|\psi_1><\psi_1|{\frac{-V_{+1}}{\sqrt 2}}|0> \nonumber \\
& \hspace{0.8cm}  +  <0|\frac{-U_{+1}}{\sqrt 2}|\psi_{-1}><\psi_{-1}|{\frac{V_{-1}}{\sqrt 2}}|0>
\end{eqnarray}
In an exactly analogous manner the y-y term reduces to:

\begin{eqnarray}
{\sum_{i} {<{S_{1n}^{y}}>}_{0i} {<{S_{1n^\prime}^{y}}>}_{i0}} = & 
  - <0|\frac{ U_{-1}}{\sqrt 2}|\psi_1><\psi_1|{\frac{V_{+1}}{\sqrt 2}}|0> \nonumber \\
&\hspace{0.35cm}  -  <0|\frac{U_{+1}}{\sqrt 2}|\psi_{-1}><\psi_{-1}|{\frac{V_{-1}}{\sqrt 2}}|0>
\end{eqnarray}
which is the same as the x-x term. Finally the z-z term is given by,

\begin{equation}
{\sum_{i} {<{S_{1n}^{z}}>}_{0i} {<{S_{1n^\prime}^{z}}>}_{i0}} =
  <0| U_{0}|\psi_0><\psi_0|V_{0}|0>
\end{equation}
Once we have proven that the x-x and y-y terms are equal the above has to be equal to 
the other two by rotational invariance. This can also be seen explicitly by application of the 
Wigner-Eckart theorem. The arguments are  the following:

\begin{itemize}
\item{The matrix element of the form $<\psi_i|\hat O|0>$ wherever it occurs must have the same
value everywhere as all the relevant Clebsch-Gordon coefficients are 1 (we are adding $J_1 = 0$ and $J_2 =1$)
and the other term that we need to evaluate is the same for all such elements
 as it does not depend on the $S_z$ values.}
\item{The matrix element of the form $<0| \hat O|\psi_i>$ even though has the same magnitude in
all the three equations  has the opposite sign in the z-z term,
this is because $<1,\pm 1;1,\mp 1|1,1;0,0>=-<1,0;1,0|1,1;0,0>$ (we have used the notation $<j_1,j_{1z};j_2,j_{2z}|j_1,j_2,j^{tot},j_z^{tot}>$).
This negates the sign difference in the right hand sides of Eqs 3.5, 3.6 and 3.7. Moreover the factor 
of $\frac{1}{2}$ in the x-x and y-y terms is also accounted for by the fact the the x-x and the 
y-y terms contain the sum of two terms each of which are equal in magnitude to the z-z term.
Thus having proved that the x-x, y-y and the z-z terms are
equal we have proved (A).}
\end{itemize}

\begin{flushleft}
{\bf{\underline{Proof of (B)}}}
\end{flushleft}
We first consider the x-y term. We have after eliminating terms using the m selection rule,
\begin{eqnarray}
{\sum_{i} {<{S_{1n}^{x}}>}_{0i} {<{S_{1n^\prime}^{y}}>}_{i0}} = & 
  + <0|\frac{ U_{-1}}{\sqrt 2}|\psi_1><\psi_1|{\frac{iV_{+1}}{\sqrt 2}}|0> \nonumber \\
& -  <0|\frac{U_{+1}}{\sqrt 2}|\psi_{-1}><\psi_{-1}|{\frac{iV_{-1}}{\sqrt 2}}|0>
\end{eqnarray}
we see that both terms again have the same magnitude (note the point about the relevant C-G coefficients
in the proof of (A)) but opposite sign so this term vanishes.

Thus y-z and the x-z term must also vanish by rotational invariance (these can be also be
 shown explicitly using arguments similar to those used in the proof of {\bf{(A)}}) .
Thus having proved the two assertions we now come to the conclusion that the effective 
Hamiltonian governing the almost decoupled spin-$\frac{1}{2}$'s (close to the ground state)
to the second order in perturbation theory is given by two-spin interactions of the form  
$\overrightarrow S_i \cdot \overrightarrow S_j$. The final form will thus look like,
{\boldmath
\begin{eqnarray}
 H_{eff} = a + \frac{J_2^2}{J_1}[c_1({\overrightarrow S}_1.{\overrightarrow S}_2 +{\overrightarrow S}_2.{\overrightarrow S}_3....) 
\nonumber \\
+c_2({\overrightarrow S}_1.{\overrightarrow S}_3+{\overrightarrow S}_2.{\overrightarrow S}_4...) 
\nonumber \\
+c_3({\overrightarrow S}_1.{\overrightarrow S}_4+{\overrightarrow S}_2.{\overrightarrow S}_5...)]   
\end{eqnarray}
}
where  $a \equiv Na_0J_1 + N{\frac{J_2^2}{J_1}} b_0 $ and $c_1$, $c_2$, $c_3$ etc (upto a factor
of $J_2^2/J_1$) are the 
coupling strengths between the nearest neighbours, next-nearest-neighbours and so on.
The first term in $a$ corresponds to the energy of the spin-1 system and the second term
comes from spin-$\frac{1}{2}$ terms of the form $\overrightarrow S_i\cdot \overrightarrow S_i$.

We note here that the $c_i$'s involve a sum over matrix elements connecting all the spin-1 excited
states with the singlet ground state (eq 3.3). Analytically calculating  this sum is difficult
as we do not have the complete information of all such states in order to calculate the required matrix 
elements. Thus we calculated the coefficients numerically. The method used is described in the following
section.

\subsection{Calculation of interaction strengths between the spin-$\frac{1}{2}$'s}

The computation of the interactions strengths was done using $\delta (\equiv {J_1}/{J_2}) = 10$ and
for a system size of 8 triangles.
This ensures that the perturbative corrections are convergent as the matrix elements calculated in 
eq 3.3 are of the order $J_2$ and for convergence of the perturbation theory this must be greater than
the difference in the energy between the ground state and the first excited state of the spin-1
system. That is known to be of the order $J_1$.

We calculated the interaction strengths $(J_2^2/J_1)c_1$, $(J_2^2/J_1)c_2$ etc as follows:

\begin{itemize}
\item{We begin by reducing all the bond strengths to the spin-$\frac{1}{2}$s to zero.
This will give us the constant $a_0$.}
\item{Now we connect the bonds with strength $J_2$ to only two of the  spin-$\frac{1}{2}$ s.
We do this successively for nearest neighours next nearest neighbours and so on.}
\item{There being only two spin-$\frac{1}{2}$ s in the system, the effective Hamiltonian
governing them to second order in perturbation will be of the form  
{\bf{A + B$\mathbf \overrightarrow S_i\cdot \overrightarrow S_j$}}}.
\item{$\overrightarrow S_i$ and $ \overrightarrow S_j$ being spin-$\frac{1}{2}$ s we know that
the ground state and the first excited states will have values $A - \frac{3}{4}B$ and $A + \frac{1}{4}B$
or vice versa. Which of these is the ground state depends on the sign of the coupling. If the coupling is 
ferromagnetic, the ground state will be a triplet and the latter will be the ground state
and else the former will be the ground state.}
\item{Thus knowing the ground state and the first excited state and thus $A$ and $B$ we get $b_0$
and the $c_i$ s.}

\end{itemize}
\begin{flushleft}
The various coefficients of eq 3.9 calculated using the above method (for a system size of 8 triangles)
turn out to be,\\
\end{flushleft}
\begin{center}
\begin{tabular}{|r l|}
\hline
{\large{\boldmath{$a_0 = -1.41712$}}}, & {\large{\boldmath{$b_0 = -0.12665$ }}}\\
{\large{\boldmath{ $ c_1 = 0.0183$}}}, & {\large{\boldmath{$c_2 = 0.1291$}}} \\
{\large{\boldmath{ $c_3 = -0.0108$}}}, & {\large{\boldmath{ $c_4 = 0.0942$}}} \\
\hline
\end{tabular} 
\end{center}

The important feature that we notice is that the {\bf{next nearest neighbour coupling $c_2$ is 
stronger than the nearest neighbour coupling $c_1$ .}}
This is the reason why the spin-$\frac{1}{2}$ -  spin-$\frac{1}{2}$ correlations between the 
next-nearest-neighbours is larger than the correlations between the nearest neighbours.

To this order in perturbation theory, the effective Hamiltonian seems to have 
the pattern AAFA (A $\rightarrow$ antiferromagnetic, F $\rightarrow$ Ferromagnetic).
One curious thing that we notice in the spin-$\frac{1}{2}$ - spin-$\frac{1}{2}$ correlation
plot  is that though $c_2$ and $c_4$ are both
positive (antiferromagnetic) the relevant  correlations seem to be opposite in sign.
The reason for this is that though both couplings are antiferromagnetic, $c_2$ larger
and thus it exercises greater control over the alignment of $\overrightarrow S_5$ 
than $c_4$. That is why though $c_4$ would dictate $\overrightarrow S_1$ and
 $\overrightarrow S_5$ to be oppositely aligned, $c_2$ being larger will win over
in trying to align $\overrightarrow S_1$ and  $\overrightarrow S_5$ in the same 
direction. This is what results in the correlations of opposite sign.

For a system size of 10 triangles, the spins \{1,3,5,7,9\} and \{2,4,6,8,10\}
because of a large $c_2$ will form two distinct frustrated systems and thus 
the correlations reported for this size will be smaller than for system sizes
which have even number of triangles (this has been checked for 8 triangles). The 
general features of the effective Hamiltonian nevertheless remain unchanged as
we can see from the plots.  

{\bf{Higher orders in perturbations theory:}} It is an interesting question to ask 
if the higher orders in perturbation theory will contribute in this case ($\delta = 10$).
Analytically the effective Hamiltonian will have a more complicated form than just the 
two-spin interaction and we cannot use the method just described 
used to calculate $c_i$ numerically.
We can try to eliminate the higher order effects by going to a higher $\delta$, say 100.
But calculating the eigenvalues corresponding to $\delta = 100$ to compute $c_i$ to required
accuracy is difficult. But if we look at the graph we see that even for $\delta=2$ we
already see the features that we have predicted for $\delta =10$ quite clearly.
This gives us an indication of the fact that the higher orders in all probability 
wont change the general features  at $\delta = 10$.

\chapter{Spin wave analysis}

In this chapter we look at the spin wave spectra obtained assuming the 
classical ground states described in Chapter 1 to be good approximations
to the quantum ground states.  

\section{Spin wave spectra in the Ferrimagnetic state}
The ferrimagnetic state which is stable for $J_2S_2>2J_1S_1$, is characterised by a 
collinear arrangement of spins with the spin-$\frac{1}{2}$'s and the spin-1's pointing
along opposite directions. Thus the spin-1's and spin-$\frac{1}{2}$'s belong to two differnt 
sublattices having net magnetic moments in opposite directions. We proceed as follows to do the spin wave
calculation for this phase, 

We write the Hamiltonian in the form,
\begin{equation}
H = H_1 + H_2
\end{equation}
where $H_1 \equiv J_1{\sum_n  {\overrightarrow S}_{1n}\cdot{\overrightarrow S}_{1(n+1)}}$,
and $H_2 \equiv J_2 \sum_n  ({\overrightarrow S}_{2n}\cdot{\overrightarrow S}_{1n} +
{\overrightarrow S}_{2n}\cdot{\overrightarrow S}_{1(n+1)})$.
\pagebreak

We now introduce the bosonic variables (we assume that the spin-1's are pointing along the 
+z direction),
\begin{eqnarray}
S_{1n}^+ = \sqrt {2S_1} a_n,  & S_{2n}^+ = \sqrt {2S_2} b_n^{\dagger} \nonumber \\
S_{1n}^- = \sqrt{2S_1} a_n^{\dagger},  & S_{2n}^- = \sqrt{2S_2} b_n  \\
S_{1n}^z  =  S_1 -  a_n^{\dagger}a_n,  & \hspace{0.65cm} S_{2n}^z = -S_2 + b_n^{\dagger} b_n \nonumber 
\end{eqnarray}
which are essentially the Holstein-Primakov transformations for the two sublattices
truncated to the lowest order (the equations for $S_z$ though are not 
approximations). And being bosonic variables 
$[a_n,a_{n \prime}^{\dagger}]=\delta_{n,n\prime}$ and $[b_n,b_{n \prime}^{\dagger}]=\delta_{n,n\prime}$.

We also introduce the bosonic variables in the reciprocal space,
\begin{eqnarray}
a_n = \frac{1}{\sqrt N} \sum_{n} a_k e^{ikn}, b_n = \frac{1}{\sqrt N} \sum_{n} b_k e^{-ik({n+\frac{1}{2})}}
\end{eqnarray}
We assume unit lattice spacing. Thus the wave vectors $\{k\}$ go from $-\pi$ to $\pi$ in units 
of ${2\pi}/{N}$ where N is the number of sites on each sublattice.

Having defined these variables we now write $H_1$ and $H_2$ in terms of these bosonic 
operators assuming the classical configuration of the spins in the ferrimagnetic state
to be a good approximation to the quantum ground state. If we look at $H_1$  we see
that the configuration of spin-1s mimics the ferromagnetic ground state of a Heisenberg
ferromagnet in one dimension.
So we can directly put down the form of $H_1$ in terms of the bosonic operators 
taking into account the fact that in the present case $J_1>0$.
It will be,
\begin{equation}
H_1 = NJ_1S_1^2 - 4J_1S_1 \sum_k \sin^2 (\frac{k}{2}) a_k^{\dagger}a_k
\end{equation}
Now using Eqs 4.2 and 4.3 we  write $H_2$ too in terms of the fourier variables.
This will have the form,
\begin{eqnarray}
H_2 =& -2N J_2 S_1 S_2 + 2 J_2 S_1 \sum_k b_k^{\dagger}b_k + 2J_2 S_2 \sum_k a_k^{\dagger}a_k  \nonumber \\
&\hspace{-1.70cm} +2J_2 \sqrt {S_1S_2}[\sum_{k}(b_k^{\dagger} a_k^{\dagger} + b_ka_k)\cos \frac{ka}{2}]
\end{eqnarray}
Finally adding up $H_1$ and $H_2$ we get the full Hamiltonian in the fourier variables to be,
\begin{eqnarray}
H    = NJ_1S_1^2 - 2NJ_2S_1S_2   \nonumber  &\\
&  \hspace{-4.0cm}+ 2J_2S_1 \sum_{k} b_k^{\dagger}b_k + \sum_k (2J_2S_2 - 4J_1S_1\sin^2 \frac{k}{2}) a_k^{\dagger}a_k   \nonumber  \\
& \hspace{-6.2cm} + \sum_{k} (2J_2 \sqrt {S_1S_2} \cos \frac{k}{2})(b_k^{\dagger}a_k^{\dagger} + b_ka_k)  
\end{eqnarray}
We note that the first term is the energy of the classical ground state.
For convenience of expression we now define,
\begin{eqnarray}
A_k = 2 J_2 S_2 - 4 J_1 S_1 \sin^2 (\frac{k}{2}), & B_k = 2J_2S_1, & C_k = 2 J_2 \sqrt {S_1S_2} \cos \frac{k}{2}
\end{eqnarray} 
($B_k$ is $2J_2S_1$ for all k). Thus we have,
\begin{equation}
H = NJ_1S_1^2 - 2NJ_2S_1S_2  
+ \sum_k A_k a_k^{\dagger}a_k +  \sum_{k} B_k b_k^{\dagger}b_k
+ \sum_{k} C_k(b_k^{\dagger}a_k^{\dagger} + b_ka_k) 
\end{equation}
We now use the Bogoliubov transformation to bring the above Hamiltonian to 
a more conveniently diagonalisable form. They are,
\begin{eqnarray}
a_k = u_k \cosh \theta_k - v_k^{\dagger} \sinh \theta_k \nonumber \\
b_k = v_k \cosh \theta_k - u_k^{\dagger} \sinh \theta_k
\end{eqnarray}
$u_k$ and $v_k$  satisfy the same bosonic commutation relations as $a$ and $b$. We for now
leave $\theta_k$ as undetermined.
Writing the Hamiltonian (4.8) in terms of the new variables $u_k$ and $v_k$,
\begin{eqnarray}
H =&  -J_1 N S_1^2 - N J_2 S_1 S_2 - \sum_k C_k \sinh {2\theta_k} + \sum (A_k + B_k)\sinh^2 \theta_k
\nonumber \\ 
 &\hspace{-1.85cm}  + \sum_k (A_k \cosh^2 \theta_k  + B_k \sinh^2 \theta_k - C_k \sinh 2 \theta_k) u_k^{\dagger}u_k
\nonumber \\
 & \hspace{-1.85cm}   + \sum_k (A_k \sinh^2 \theta_k  + B_k \cosh^2 \theta_k - C_k \sinh 2 \theta_k) v_k^{\dagger}v_k
\nonumber \\
 & \hspace{-1.05cm}  + \sum_k \lbrack -(\frac{A_k}{2} + \frac{B_k}{2}) \sinh 2 \theta_k + C_k \cosh 2 \theta_k \rbrack 
( u_k v_k + u_k^{\dagger} v_k^{\dagger})
\end{eqnarray}   
The last term suggests that we choose $\theta_k$ according to the following definition:
\begin{equation}
\tanh 2 \theta_k = \frac {2 C_k}{A_k + B_k}
\end{equation}
This reduces the Hamiltonian apart from constant numbers (which we can ignore
by setting the zero of energy appropriately) to a form in which all the operators are 
of the form $\hat O^{\dagger} \hat O$.

Thus the Hamiltonian finally looks like,
\begin{eqnarray}
H = \sum_k (A_k \cosh^2 \theta_k  + B_k \sinh^2 \theta_k - C_k \sinh 2 \theta_k) u_k^{\dagger}u_k
\nonumber \\
    + \sum_k (A_k \sinh^2 \theta_k  + B_k \cosh^2 \theta_k - C_k \sinh 2 \theta_k) v_k^{\dagger}v_k
\end{eqnarray}
where we have dropped the constants. So we have two modes given by,
\begin{eqnarray}
\omega_{-} =  A_k \cosh^2 \theta_k  + B_k \sinh^2 \theta_k - C_k \sinh 2 \theta_k,
\nonumber \\
\omega_{+} =   A_k \sinh^2 \theta_k  + B_k \cosh^2 \theta_k - C_k \sinh 2 \theta_k
\end{eqnarray} 
The mode $\omega_{-}$ is gapless as $k \rightarrow 0$ and is also dispersionless when $J_2S_2 = 2J_1S_1$.
The mode $\omega_{+}$ is always gapped as shown in the figure below:

\begin{center}
\epsfxsize=9cm \epsfysize=9cm \epsfbox{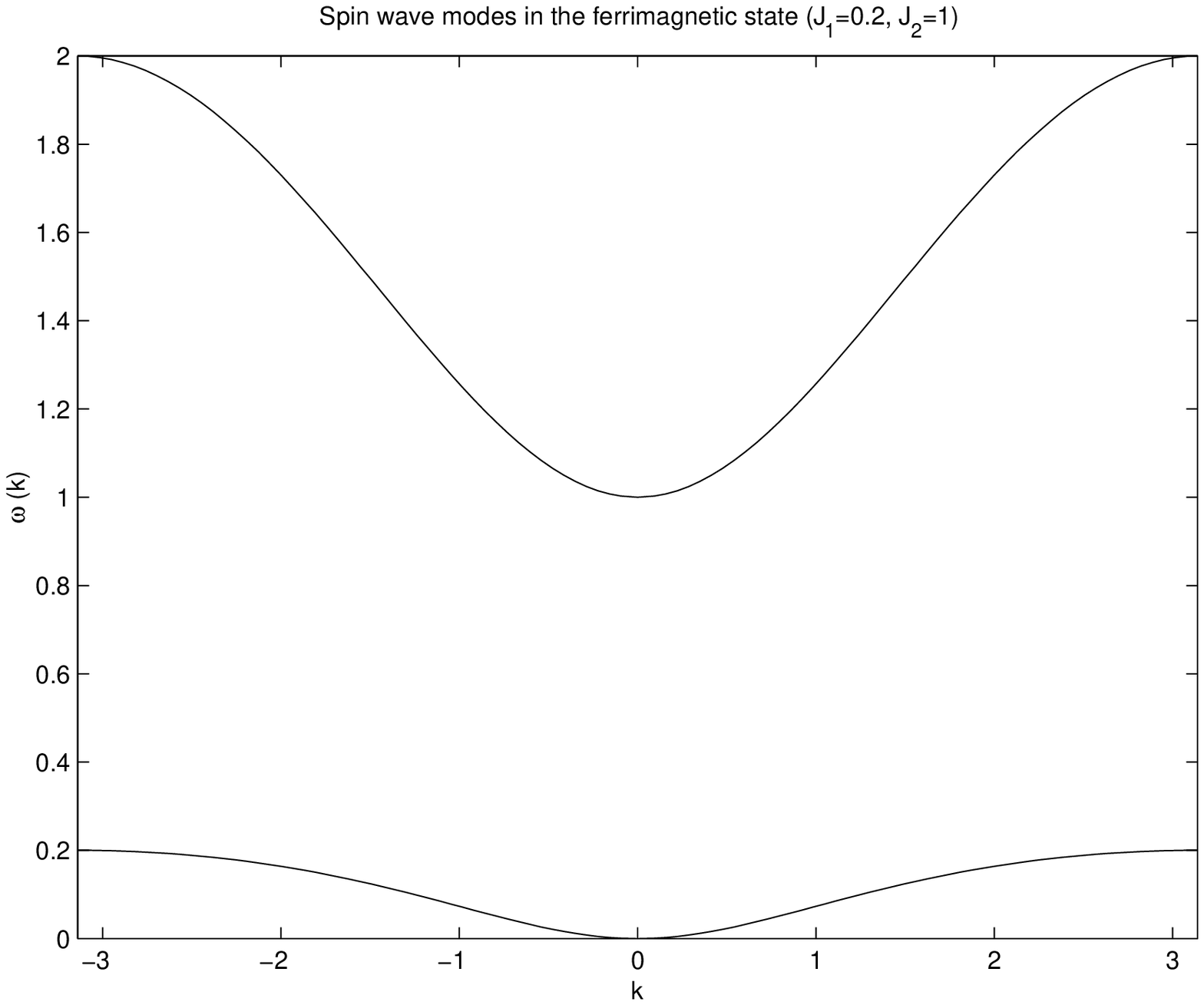}
\end{center}

\section{Spin wave spectra in the Spiral state}
We now investigate the spin wave spectra in the spiral state. 
This phase is stable for $J_2S_2<2J_1S_1$. We assume a coplanar configuration
of spins in the system as the classical ground state as shown in the figure of the same
in Chapter 1. The basic unit of this ground state ( which keeps repeating) schematically looks as
follows:

\begin{flushleft}
\epsfxsize=17cm  \epsfysize=7cm \epsfbox{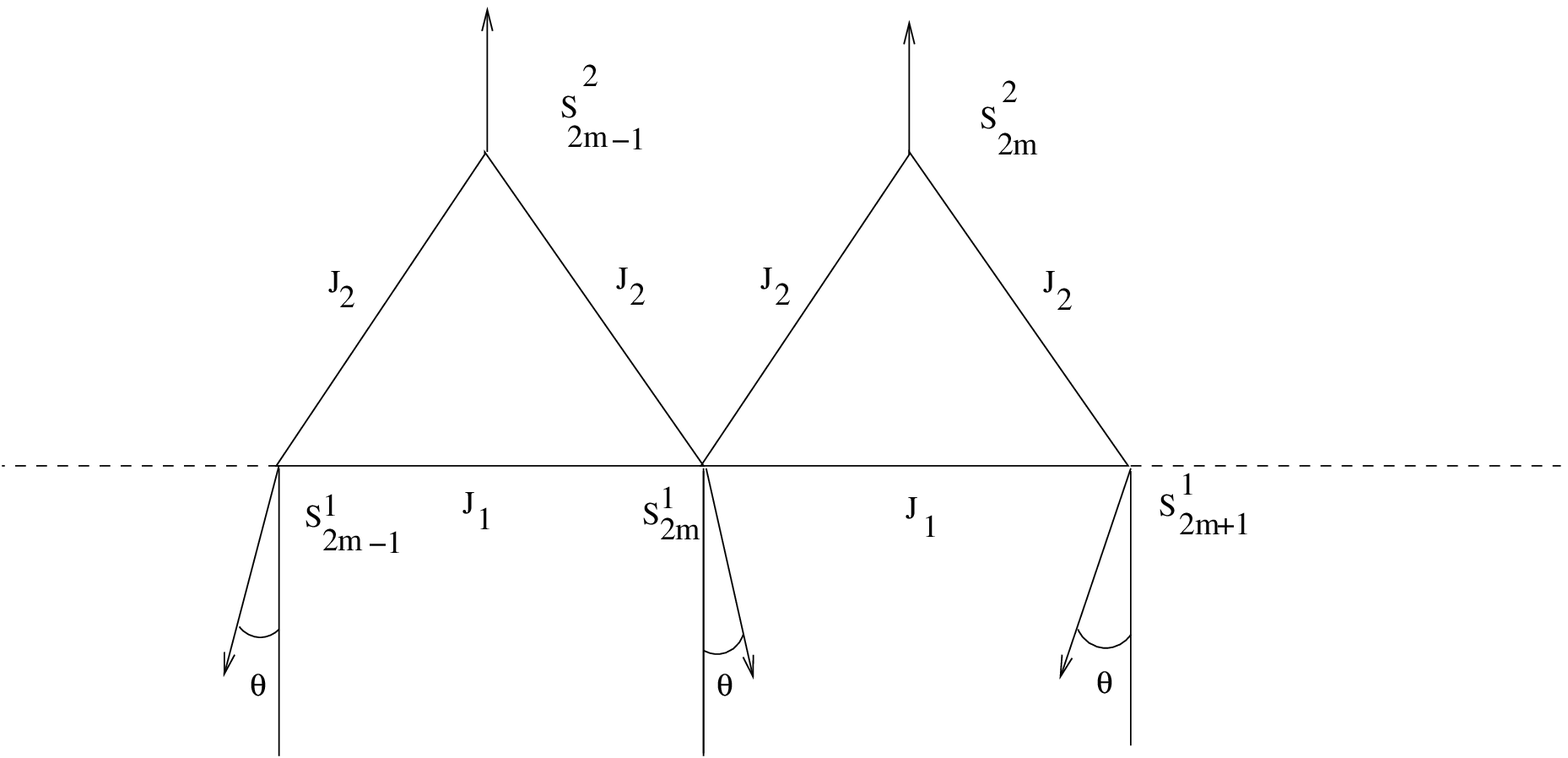}
\end{flushleft}
where $\cos \theta \equiv {J_2S_2}/{2J_1S_1}$ and $S^1_n$ and $S^2_n$ denote spin-1's
and spin-$\frac{1}{2}$'s respectively and the subscripts are the site indices. In order to 
do valid spin-wave calculations with this classical configuration, for the spin-1s on the 
baseline we must choose(for each spin-1 separately)
the +z direction to be along (or opposite to)
the classically expected direction of the spin vector. In doing so,
we obtain the following transformation equations for the components 
of spin vectors (and thus for the corresponding operators) corresponding 
to the spin-1s,
\begin{eqnarray}
S^{1x}_{2m} = S^{1x\prime}_{2m}\cos \theta - S^{1z\prime}_{2m} \sin \theta,
& S^{1x}_{2m-1} =  S^{1x\prime}_{2m-1}\cos \theta + S^{1z\prime}_{2m-1} \sin \theta \nonumber \\
S^{1z}_{2m} = S^{1x\prime}_{2m}\sin \theta + S^{1z\prime}_{2m} \cos \theta,
& S^{1z}_{2m-1} = -S^{1x\prime}_{2m-1}\sin \theta + S^{1z\prime}_{2m-1} \cos \theta \nonumber \\
S^{1y}_{2m} = S^{1y\prime}_{2m}, & S^{1y}_{2m-1} = S^{1y\prime}_{2m-1}
\end{eqnarray}

Here we have chosen the direction in which spin-$\frac{1}{2}$'s are aligned to be  the +z direction and
the +y direction is into the page. The primed coordinates are obtained by
rotation about the $y$-axis by an angle $\theta$.
Because of the different classical orientations of the spin-1's on the 
even and odd numbered sites  we see that the transformation equations 
for them  are different ($\theta$ has changed sign). We also note that 
the classical spin-1 vectors are along the $-z\prime$ directions at each site.
We can now proceed with the spin wave calculation as the approximation used 
(that the deviation of $S_z$ from $S_1$ is small) is valid if we use the primed coordinates
for the spin-1's.

As opposed to the ferrimagnetic case here we have to define three categories of
bosonic variables. They are 
\begin{flushleft}
\begin{eqnarray}
S_{2m}^{1+} = \sqrt {2S_1} b_{2m},&S_{2m-1}^{1+} = \sqrt {2S_1} d_{2m-1},  & S_{n}^{2+} = \sqrt {2S_2} c_n^{\dagger} \nonumber \\
S_{2m}^{1-} = \sqrt{2S_1} b_{2m}^{\dagger},&S_{2m-1}^{1-} = \sqrt{2S_1} d_{2m-1}^{\dagger},  & S_{n}^{2-} = \sqrt{2S_2} c_n \nonumber \\
S_{2m}^{z\prime} = -S_1 +  b_{2m}^{\dagger}b_{2m},&S_{2m-1}^{z\prime} = -S_1 +  d_{2m-1}^{\dagger}d_{2m-1},  & S_{n}^{2z} = S_2 - c_n^{\dagger} c_n 
\end{eqnarray}
\end{flushleft}
Where the $b$'s $d$'s are the bosonic variables for the spin-1's at 
the even and the odd numbered sites respectively (in the above expressions,
$S^{1+}_{2m}$ etc have been defined in terms of the primed components). The index
$n$ runs over all the spin-$\frac{1}{2}$ sites. From here the calculation proceeds
in the following stages:
\begin{enumerate}
\item{We use Eq. 4.14 and Eq. 4.15 to write the Hamiltonian (4.1) in terms of the three bosonic
variables in the unprimed coordinates.}
\item{The first order terms in $b_{2m}$'s,  $d_{2m-1}$'s and $c_n$'s vanish. We thus keep 
the terms to the second order in the above variables as the approximation to the Hamiltonian.}
\item{At this point the Hamiltonian doesn't have a simple form which can be diagonalised
using the Bogoliubov transformation used in the ferrimagnetic case. We use another method
to find the spectrum here. We define a new set of canonically conjugate 
variables using the following equations,
\large{
\begin{eqnarray}
\sqrt 2 b_{2m} = {q_{b2m}+ip_{b2m}}, & \sqrt 2 b_{2m}^{\dagger} = {q_{b2m}-ip_{b2m}} \nonumber \\
\sqrt 2 d_{2m-1} = {q_{d(2m-1)}+ip_{d(2m-1)}}, & \sqrt 2 d_{2m-1}^{\dagger} = {q_{d(2m-1)}-ip_{d(2m-1)}} \nonumber \\
\sqrt 2 c_n      = {q_{cn} + ip_{cn}}, & \sqrt 2 c_n^{\dagger} = {q_{cn} - ip_{cn}}
\end{eqnarray}
}
}
\item{Now we write the Hamiltonian obtained in step 2 in terms of these operators.
In terms of these operators we find a couple of properties of the Hamiltonian.
The q's and the p's don't couple in any of the terms. Secondly  $q_{b2m}$'s and $q_{d(2m-1)}$'s
and the corresponding momenta occur completely symmetrically in the Hamiltonian.There is no way 
to choose one over the other. 
So instead of having two different variables for the odd and even numbered spin-1 sites 
we can express the Hamiltonian using just one set of variables for all spin-1's.
We call that set $Q_{1n}$ and $P_{1n}$ where n runs over all spin-1's. For convenience 
of expression we now call $q_{cn}, p_{cn}$  $Q_{2n}$ and $P_{2n}$ respectively and n here
runs over all spin-$\frac{1}{2}$'s.}

\item{After all the above simplifications the Hamiltonian finally looks like (omitting
the constant term),
\begin{eqnarray}
H = & \hspace{-3.8cm}  \sum_n A (Q_{1n}^2 + P_{1n}^2) + \sum_n B(Q_{2n}^2 + P_{2n}^2 ) \nonumber \\
 &  + \sum_n  C{\cos {2\theta}}\hspace{0.25cm} {Q_{1n}Q_{1(n+1)}}  + \sum_n D \cos \theta \hspace{0.25cm} Q_{2n}(Q_{1n} + Q_{1(n+1)}) \nonumber \\
& \hspace{-2.4cm} + \sum_n C \hspace{0.15cm} P_{1n}P_{1(n+1)}   - \sum_n D \hspace{0.15cm} P_{2n} (P_{1n} + P_{1(n+1)})
\end{eqnarray} 
where, $A = J_2 S_2 \cos \theta - J_1 S_1 \cos 2 \theta = J_1S_1$ (using the definition of $\cos \theta$),
$B = J_2 S_1 \cos \theta$, $C = J_1 S_1$ and $ D = J_2 \sqrt {S_1S_2}$.  
}
\end{enumerate}
Clearly this is the Hamiltonian for coupled linear harmonic oscillators 
 with nearest neighbour 
and next-to-nearest neighbour interactions. This tells us that for the purpose
of finding the eigenfrequencies  we can use the classical equations of motion.
This is because the eigenfrequencies of any such system on quantisation turn out
to be the same as the classical ones. Thus using the Hamilton's equations of motion
and the trial solutions,
\begin{eqnarray}
Q_{1n} = \epsilon_{1q} \exp i(kn -\omega t), & P_{1n} = \epsilon_{1p} \exp i(kn -\omega t) \nonumber \\
Q_{2n} = \epsilon_{2q} \exp i(kn -\omega t), & P_{2n} = \epsilon_{2p} \exp i(kn -\omega t)
\end{eqnarray} 
(where $k \hspace{0.2cm} \epsilon \hspace{0.2cm} \{-\pi,\pi\}$) we get the following matrix equations,
\begin{eqnarray}
{\mathbf{ \dot P_n = - A^\prime Q_n}} \nonumber \\
{\mathbf{ \dot Q_n =  B^\prime P_n}} 
\end{eqnarray}

where,  
\begin{displaymath}
\mathbf{ P_n} =
\left [ \begin{array}{c}
P_{1n}  \\
P_{2n} 
\end{array} \right ] \hspace{0.2cm} and \hspace{0.2cm}
\mathbf{ Q_n} =
\left [ \begin{array}{c}
Q_{1n}  \\
Q_{2n}\\
\end{array} \right ]
\end{displaymath}

$\mathbf A^{\prime}$ and $\mathbf B^{\prime}$ are calculated to be,

\begin{displaymath}
\mathbf{ A^{\prime}} = 
\left ( \begin{array}{cc}
2A + 2C\cos 2\theta \cos k & 2D e^{-\frac{ik}{2}}\cos \theta \cos \frac{k}{2} \\
2D e^{\frac{ik}{2}}\cos \theta \cos \frac{k}{2} & 2B \\
\end{array} \right )
\end{displaymath}

\begin{displaymath}
\mathbf{ B^{\prime}} =
\left ( \begin{array}{cc}
2A + 2C \cos k & 2D e^{-\frac{ik}{2}} \cos \frac{k}{2} \\
2D e^{\frac{ik}{2}} \cos \frac{k}{2} & 2B \\
\end{array} \right )
\end{displaymath}

Clearly the eigenvalues of the matrix $\mathbf{B^{\prime}A^{\prime}}$ will 
give us the squares of the eigenfrequencies. The following can be easily
shown using the definitions of A,B etc.
\begin{eqnarray}
Det B^{\prime} = 0, & Det A^{\prime} \neq 0
\end{eqnarray}
Both of these together clearly imply that {\bf{one of the eigenvalues of
 $\mathbf{B^{\prime}A^{\prime}}$ will be 0}}. Since the trace of
$\mathbf{B^{\prime}A^{\prime}}$ will be the sum of its eigenvalues, the other mode 
can be calculated by calculating the trace of $\mathbf{B^{\prime}A^{\prime}}$ as one 
of the eigenvalues has already been determined to be 0. After taking the trace 
of  $\mathbf{B^{\prime}A^{\prime}}$, we find the other mode to be given by
\begin{equation}
\omega^2 = 4 J_2^2 [ {(S_1 \cos \theta - S_2 \cos^2 (\frac{k}{2}))}^2 + (\frac{S_2^2 }{4}) \tan^2 \theta \sin^2 k ].
\end{equation} 

The above mode has a minimum at $k=0$ ($\omega (k=0) = 2 J_2 S_2 |1 - \frac{J_2}{2J_1}|$)
which vanishes at $\frac{J_2}{J_1}=2$.
At this point the nonvanishing mode looks as shown in the figure:
\begin{center}
\epsfxsize=9.0cm \epsfysize=9.0cm \epsfbox{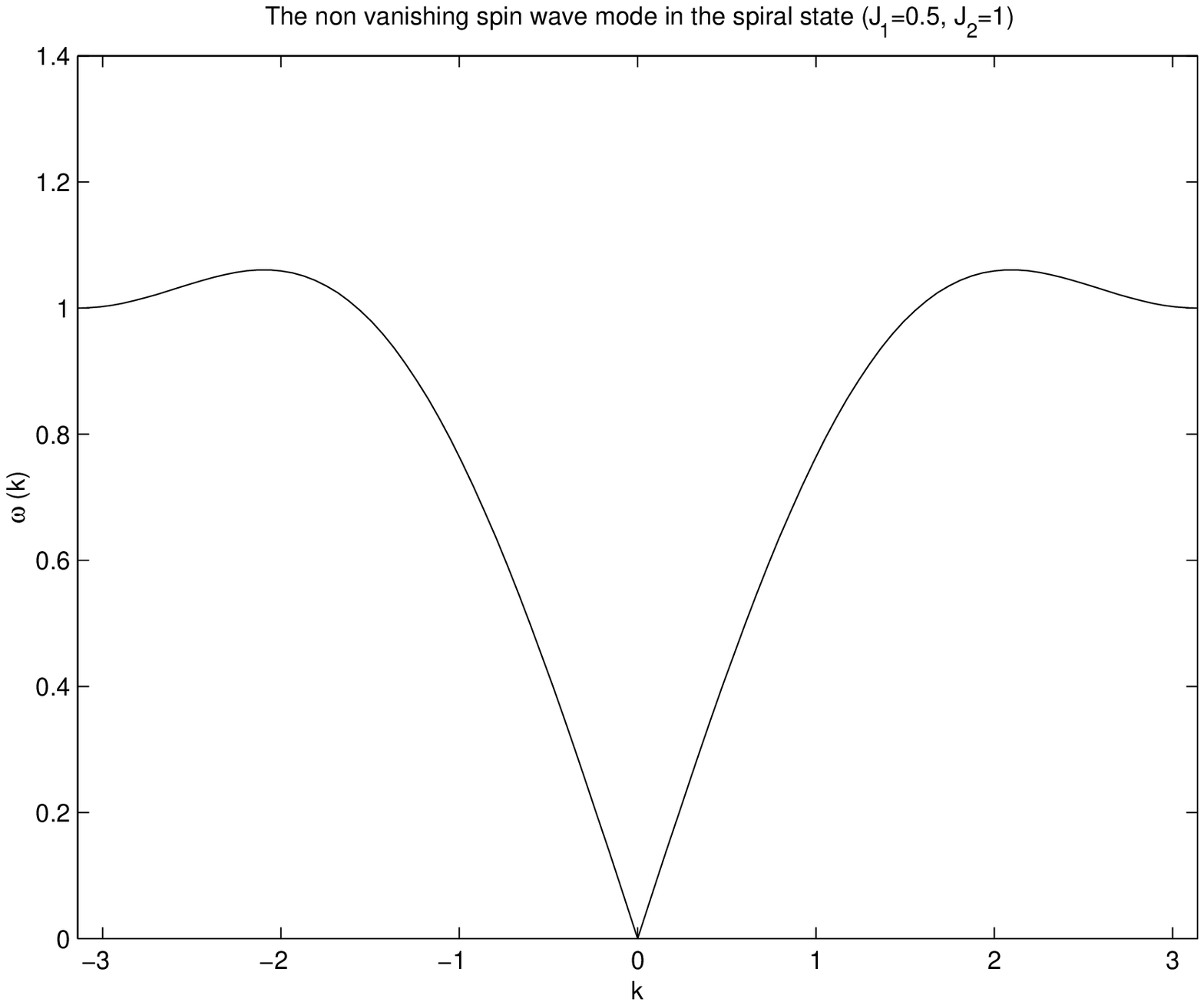}
\end{center}

Thus the spiral state spin wave spectra 
consists of  two modes one of which is dispersionless and gapless and the other 
gapped with the gap vanishing at $\delta = 0.5$. Notably $\delta=0.5$ is indeed 
one of the points where the system has been observed to be gapless numerically.
Spin wave analysis doesn't give an indication of the point $\delta=1.0$ being gapless.

\section{{Conclusions}}

To conclude, we have studied numerically and analytically, 
 a two-spin (1 and 1/2) variant of the antiferromagnetic sawtooth 
lattice. Interesting features brought to light by this study are the 
following:

\begin{itemize}
\item{The system seems to be gapless at two points in the phase diagram.
Thus there seem to be more than the two expected phases from the classical 
analysis.
Moreover none of these points correspond to the classically expected point
for the transition from the ferrimagnetic to the spiral phase.
Spin wave analysis does give some indication of the point $\delta =0.25$
being gapless but the points which are actually found to be gapless are 
$\delta=0.5$ and  $\delta=1.0$. This may be due to the small values of spins
because of which spin wave analysis may not be very accurate. The properties of
the system at $\delta=0.5$ and $\delta=1.0$ are unclear as of now.

}

\item{In the large $\delta$ limit, though the other correlations behave as expected,
the correlations between the spin-$\frac{1}{2}$'s have the curious feature that 
the strongest coupled spins are the next-nearest-neighbours. This we have been
able to explain using perturbation theory analysis of the model.}
\end{itemize}

The study of the nature of the quantum phases at the points $\delta=0.5$ and
$\delta=1.0$ is an interesting direction in which further work on this model can proceed.
Numerically one can try to go the larger system sizes using DMRG to eliminate the finite
size effects and thus better approximate the thermodynamic limit. The aspect of
the problem we haven't touched on at all is the effect of a magnetic field and the 
thermodynamics of the system. This is a another direction of study which can lead to a better understanding of this
model.

\addtocontents{toc}{{\bf  References \\}}


\begin{thebibliography}{99}
\bibitem{bss}  D. Sen, B. S. Shastry, R. E. Walstedt and R. Cava, Phys. Rev. B {\bf{53}}, 6401 (1996)
\bibitem{blun} S. A. Blundell and M. D. Nunez-Regueiro, cond-mat/0204405
\bibitem{rud} I. Rudra, D. Sen and  S. Ramashesha, cond-mat/0210122
\bibitem{skp} Swapan. K. Pati, S. Ramasesha and  D. Sen, Phys. Rev. B {\bf{55}}, 8894 (1997)
\bibitem{ivanov1} N. B. Ivanov and J. Richter, Phys. Rev. B {\bf{63}}, 144429 (2001)
\bibitem{ivanov2} N. B. Ivanov, Phys. Rev. B {\bf{62}}, 3271 (2000)
\bibitem{ivanov3} J. Richter, U. Schollw{$\ddot{o}$}ck and  N. B. Ivanov, Physica B {\bf{281 \& 282}}, 845 (2000)
\bibitem{ivanov4} N. B. Ivanov, J. Richter and  U. Schollw{$\ddot{o}$}ck, Phys. Rev. B {\bf{58}}, 456 (1998)
\bibitem{kahn1} O. Kahn, Y. Pei and Y. Journax, {\it{Inorganic Materials}}, John Wiley \& Sons Ltd, New York, 59-114  (1992)
\bibitem{kahn2} O. Kahn, Molecular Magnetism, VCH, New York (1993).
\bibitem{kahn3} Y. Pei, M. Verdaguer, O. Kahn, J. Sletten and J. P. Renard, Inorg. Chem. {\bf{26}}, 138 (1987)
\bibitem{gleizes} A. Gleizes and M. Verdaguer, J. Am. Chem. Soc. {\bf{106}}, 3727 (1984)
\bibitem{huse} S. R. White and D. K. Huse, Phys. Rev. B {\bf{48}}, 3844 (1993)
\bibitem{cullum} J. K. Cullum and R. A. Willoughby, {\it{Lanczos Algorithms for Large Symmetric Eigenvalue Computations}},
Birkh{$\ddot{a}$}user (1985)
\end{thebibliography}
\end{document}